\documentclass[11pt]{article}
\usepackage{amsmath, amssymb}
\usepackage{slashed}
\usepackage{verbatim}
\usepackage{latexsym}
\usepackage{euscript}
\usepackage{amsfonts}
\usepackage{verbatim}
\usepackage[]{array}
\usepackage[]{mathrsfs}
\usepackage{graphicx}
\usepackage{amsmath}
\usepackage{amssymb}
\usepackage{amsxtra}
\usepackage{color}
\def\be{\begin{eqnarray}}
\def\ee{\end{eqnarray}}
\def\0{\nonumber}

\def\tr{{\rm tr}}
\def\Tr{{\rm Tr}}

\def\bfD{{\bf D}}
\def\bfG{{\bf G}}
\def\bfD{{\bf D}}
\def\bfd{{\bf d}}
\def\bfh{{\bf h}}
\usepackage{slashed}
\usepackage{verbatim}
\usepackage{latexsym}
\usepackage{euscript}

\newcommand\EW{\EuScript{W}}
\newcommand\EV{\EuScript{V}}

\newcommand\EF{\EuScript{F}}

\newcommand\EA{\EuScript{A}}
\newcommand\EC{\EuScript{C}}
\normalfont\large
\begin{document}
\begin{flushright}
SISSA/48/2018/FISI\\ZTF-EP-18-05\\
arXiv:1811.04847
\end{flushright}
\vskip 2cm
\begin{center}

{\LARGE {\bf HS in flat spacetime. The effective action method}}

\vskip 1cm

{\large  L.~Bonora$^{a}$, M.~Cvitan$^{b}$, P.~Dominis
Prester$^{c}$, S.~Giaccari$^{d}$,
T.~$\rm \check{S}$temberga$^{b}$
\\\textit{${}^{a}$ International School for Advanced Studies (SISSA),\\Via
Bonomea 265, 34136 Trieste, Italy, and INFN, Sezione di
Trieste \\ }
\textit{${}^{b}$ Department of Physics, Faculty of Science, University of
Zagreb, Bijeni\v{c}ka cesta 32,
10000 Zagreb, Croatia \\}
\textit{${}^{c}$ Department of Physics, University of Rijeka,\\
Radmile Matej\v{c}i\'{c} 2, 51000 Rijeka, Croatia\\}
\textit{${}^{d}$ Department of Sciences,
Holon Institute of Technology (HIT),\\
52 Golomb St., Holon 5810201, Israel}
}
\vskip 1cm

\end{center}

 \vskip 1cm {\bf Abstract.}
 This is the first paper in a series of three dealing with HS theories in flat
spacetime.
It is divided in three parts. The first part is an elaboration on the method of
effective action, initiated in a previous paper. We study the properties of
correlators
of currents in the free fermion coupled to external higher spin (HS) potentials,
and develop techniques for their explicit
calculation. In particular we show how they can be calculated via ordinary
Feynman
diagram techniques. We also introduce the concept of {\it curved} $L_\infty$
algebra and show how it can be realized in the context of the fermion model. In
part II we
compare the results of the scalar model and
those of the fermion model (coupled to HS fields). We show that the HS field
formulation 
coming from the scalar model is the `square' of  
the one ensuing from the fermion model.  Finally, in part III, we analyze the
possible obstructions that one may meet  in constructing the effective action:
these are the analogs of anomalies in
ordinary gauge theories. We provide explicit and compact formulas of the latter.
\vskip 1cm
\begin{center}

{\tt Email: bonora@sissa.it,mcvitan@phy.hr,
pprester@phy.uniri.hr,stefanog@hit.ac.il, tstember@phy.hr}

\end{center}

\eject

\section{Introduction}
\label{sec:intro}

This paper and the following ones are devoted to the construction and analysis
of massless field models with infinite many fields of increasing spins in flat
spacetime. 
The construction of interacting relativistic quantum field theories with
fundamental
fields/particles of spin higher then two (higher spin, or HS, for short) in a
flat background is a longstanding problem of theoretical physics. On the one
hand there are theoretical motivations for searching such theories.
On the wake of (super)string theories, it has been known that infinite towers of
higher spin fields can soften the UV behavior in a drastic way. The AdS/CFT
correspondence offers another example in which the description of an ordinary
conformal field theory can be improved by the use of a dual string theory. There
are also arguments suggesting that when gravity is involved an infinite number
of fields with increasing spin is necessary in order to avoid conflicts with
causality, \cite{Maldacena}.

On the other hand, there are some dissuasive elements in the story. Even if we
ignore the experimental evidence against the existence of elementary
HS particles, there seem to be theoretical obstructions: there are in the
literature several no-go
theorems prohibiting, under some rather standard assumptions, the existence of
massless HS particles in a flat spacetime. It is not strange then that
the only HS theories in $d\ge4$ dimensions constructed so far are in
the (A)dS background, through the Vasiliev program \cite{Vasiliev}. 
Studies of the massless HS
theories in flat spacetime clearly suggest that such theories must violate some
of the usual QFT assumptions, which are
assumed in no-go theorems: either locality (see, in particular \cite{FS}), or 
minimal coupling to gravity, or
finiteness of the number of particles with the mass below any finite energy
scale (infinite
tower of HS fields are necessary, see \cite{BBvD}).

We know that in 
the case of free HS theories,\cite{Fronsdal}, the constructions using Wigner and 
Bargmann-Wigner
methods goes the same way as for lower spin fields/particles, and that free
field theories contain infinite towers of conserved currents of all spins. These
two properties are normally starting points in constructing theories with local
symmetries and their interactions with matter fields, well known examples being
Maxwell/Yang-Mills gauge field theories and GR. Understandably this is the
reason why 
most attempts to construct HS theories have moved along these same lines.
Here, though,  we would like to pursue another interesting avenue for exploring
HS theories,
opened by papers 
\cite{Bekaert:2009ud,Bekaert:2010ky,BCDGPS}. In
\cite{Bekaert:2009ud,Bekaert:2010ky} it was shown that the Klein-Gordon field
(the matter field) linearly coupled to the infinite tower of HS background
potentials represented by symmetric tensor fields (metric-like formalism)
$h_{(s)}^{\mu_1\ldots\mu_s}(x)$
\be \label{mkglch}
S[\varphi,h] = S_0[\varphi] + \sum_{s=0}^\infty  \int
d^dx\,\frac 1{s!}
J^{(s)}_{\mu_1\ldots\mu_s}(x)\,
 h_{(s)}^{\mu_1\ldots\mu_s}(x)
\ee 
where $S_0[\varphi]$ is the action for the free (massive) KG field and
$J^{(s)}_{\mu_1\ldots\mu_s}$ is the spin-$s$ current which is conserved in the
free KG theory, is symmetric under local HS transformations. If one chooses for
the
currents the so-called simple form
{
\be
J^{(s)}_{\mu_1\ldots\mu_s}(x) = \frac{i^{s-2}}{2^s}\, \varphi(x)^\ast\!
\stackrel{\leftrightarrow}{\partial}_{\mu_1} 
\cdots \stackrel{\leftrightarrow}{\partial}_{\mu_s}\! \varphi(x)
\ee
}
then the HS symmetry has also a particularly simple form when the HS potentials
are
represented in a $2d$ dimensional {\it phase space}
$(x,u)$\footnote{Throughout the paper the position in the phase space are
denoted by couples of letters $(x,u),
(y,v),(z,t),(w,r)$, the first 
letter refers to the space-time coordinate and the second the 
worldline particle momentum. 
The letters $k,p,q$ will be reserved to the momenta of the
(Fourier-transformed) physical amplitudes.} . In this case the HS gauge
transformation of the
HS master field potential\footnote{In order to distinguish them from ordinary
spacetime fields, 
at times we will  refer to fields like $h(x,u)$ and $J(x,u)$, and  $h_a(x,u)$ 
and $J_a(x,u)$ below, as {\it master fields}.} is
{
\be \label{delh2}
\delta_{\eta} h(x,u) = (u\!\cdot\! \partial_x)\eta (x,u)
 -\frac  i2\, [ h(x,u) \stackrel{\star}{,} \eta (x,u)] 
\ee
}
where $\star$ is the Moyal product and $[\;\stackrel{\star}{,}\;]$ is the
commutator defined by the Moyal product. The HS master field potential $h(x,u)$ 
is defined by the
Taylor
expansion around $u=0$
\be \label{hmte0}
h(x,u) = \sum_{s=0}^\infty \frac{1}{s!}\, h_{(s)}^{\mu_1\cdots\mu_s}(x)\,
u_{\mu_1} \cdots u_{\mu_s}
\ee
and {${\eta}(x,u)$} is an arbitrary phase space parameter regular around
$u=0$
{
\be \label{epste}
{\eta}(x,u) = \sum_{s=1}^\infty \frac{1}{(s-1)!}\,
{\eta}^{\mu_1\cdots\mu_{s-1}}(x)\, u_{\mu_1} \cdots u_{\mu_{s-1}}
\ee
}
where {${\eta}^{\mu_1\cdots\mu_{s-1}}(x)$} are arbitrary spacetime
tensor fields parametrizing local HS transformations. By using (\ref{hmte0}) and
(\ref{epste}) in (\ref{delh2}) one gets that HS transformations of spacetime HS
fields are of the form
{
\be
\delta_{{\eta}(x,u)} h_{(s)}^{\mu_1\cdots\mu_s}(x)
 = s\, \partial^{(\mu_1} {\eta}^{\mu_2\cdots\mu_s)}(x) +
\mbox{(non-linear terms)}
\ee
}
The linear part has the standard form of linearized unrestricted HS
transformations for $s\ge1$.
 
 As was observed in \cite{Bekaert:2009ud,Bekaert:2010ky} the HS gauge
transformation
(\ref{delh2}) obeys Lie-algebra type commutator\footnote{Using different forms
for the conserved currents is equivalent to field redefinitions, which would
change the form of HS transformation (\ref{delh2}). Still, the Lie-algebra
structure
(\ref{lieam}) would of course be the same.}
{
\be \label{lieam}
[ \delta_{{{\eta}_1}} , \delta_{{{\eta}_2}} ]
 = \delta_{i [{{\eta}}_1 \stackrel{\star}{,} {{\eta}}_2]}
\ee
}

While there is simplicity and elegance in the relations (\ref{delh2}) and
(\ref{lieam}) and obvious similarity with the analogous Yang-Mills
transformations,
there are also some annoying aspects. First of all, we know
already from electrodynamics (pure spin-1 example) that coupling of KG field to
spin-1 gauge field contains also quadratic (seagull) terms in the action, which
are not present in \eqref{mkglch}. For this reason it was assumed in
\cite{Bekaert:2009ud}
that (\ref{mkglch}) is just a linearization, and that a consistently coupled
theory contains also non-linear couplings. If so, this could in principle change
the expression for HS transformations and the Lie-structure formula
(\ref{lieam}).
Moreover, the necessity of a spin-0 field in the tower of HS fields, which
couples to the mass term 
$\varphi^\ast \varphi$  (not a conserved current in the standard sense), at
first sight may look strange. However, it was noted in \cite{Bekaert:2010ky},
that the form of the HS transformations suggests that the spin-0 HS field should
be
treated as a composite field, which means that it should provide non-linear
couplings of HS fields with spin $s\ge1$ with matter. Anyway, this compositeness
drastically complicates the HS structure.

Recently, in \cite{BCDGPS} it was shown that the symmetries observed in
\cite{Bekaert:2009ud,Bekaert:2010ky} extend to other matter fields, in
particular to the Dirac field $\psi$ linearly coupled to the infinite tower of
HS background fields $h_{(s)}^{a\mu_1\ldots\mu_{s-1}}(x)$
\be \label{mdlch}
S[\psi,h] = S_0[\psi] + \sum_{s=1}^\infty\int d^dx\,
J^{(s)}_{a\mu_1\ldots\mu_{s-1}}(x)\,
 h_{(s)}^{a\mu_1\ldots\mu_{s-1}}(x)
\ee
where $S_0[\psi]$ is the action for the free Dirac field, and
$J^{(s)}_{a\mu_1\ldots\mu_{s-1}}(x)$ is the spin-$s$ current which is conserved
for the free Dirac field. If we take a simple form for the matter currents
\be
J^{(s)}_{a\mu_1\ldots\mu_{s-1}}(x) =
\frac{{(-i)}^{s-1}}{2^{s-1}(s-1)!}\, \bar{\psi}(x)
\gamma_{a}\!
 \stackrel{\leftrightarrow}{\partial}_{\mu_1} \cdots
\stackrel{\leftrightarrow}{\partial}_{\mu_{s-1}}\!\! \psi(x)
\ee
then the action (\ref{mdlch}) is invariant under the HS gauge transformations
defined by the variations
\be \label{hsvha0}
\delta_\varepsilon h_a(x,u) = \partial_a^x \varepsilon(x,u) - i [h_a(x,u)
\stackrel{\star}{,} \varepsilon(x,u)] 
\ee
where now
\be \label{hsvte0}
h_a(x,u) = \sum_{s=1}^\infty \frac{1}{(s-1)!}
h_a^{(s)a\mu_1\cdots\mu_{s-1}}(x)\, u_{\mu_1} \cdots u_{\mu_{s-1}}
\ee
The HS transformations of spacetime HS potential fields has the form
\be \label{havar}
\delta_\varepsilon h_a^{(s)\mu_1\cdots\mu_{s-1}}(x)
 = \partial_{a} \varepsilon_{(s)}^{\mu_1\cdots\mu_{s-1}}(x) + \mbox{(non-linear
terms)}
\ee

Several things deserve to be emphasized: (1) The HS field is different from the
one coupled to the KG field (because HS transformations are different). (2) The
form
of the HS variation (\ref{hsvha0}) suggests that the coupled spacetime HS fields
are symmetric only in $(s-1)$ indices, one index (denoted by latin letter)
standing out. It is usual to call such representations for HS fields {\it
frame-like.}
(3) There is no spin-0 field present, which implies that the coupling is purely
linear. (4) The HS transformation (\ref{hsvha0}) respects (\ref{lieam}){, 
with ${\eta}(x,u)$ substituted with $\varepsilon(x,u)$}.

Our idea here is to take the universality of the HS structure as a sign that the
HS symmetries obtained by linearly coupling HS fields to matter may be exact.
This assumption is supported by an observation from \cite{BCDGPS} that actions
for HS fields invariant under HS transformation (\ref{hsvha0}) have an
$L_\infty$
symmetry\footnote{{For $L_\infty$ see 
\cite{HZ,Zwiebach,Stasheff,Lada1,Lada2,Lada,Barnich, FLS}.}}, a property 
shared by many consistent theories. In the following we
will take (\ref{hsvha0}) as an exact HS gauge transformation,  and the
frame-like
approach as a starting point of our construction. We will show that 
(\ref{delh2}) is related to a {\it metric-like formulation}, the latter is a
composite
version of the frame-like one, thus clarifying the question of the multiplicity
of HS field representations. 

For the sake of practicality we split our  exposition in three papers. The
first, this paper, is tied to the idea of effective action. We improve the
analysis initiated in  \cite{BCDGPS}, by studying the properties of correlators
of currents in the fermion model and develop techniques for their explicit
calculation. In particular we show how they can be calculated via Feynman
diagram techniques. We also introduce the concept of {\it curved} $L_\infty$
algebra
and show how it can be realized in the context of the fermion model. Then we
tackle the problem of the relation between the outputs of the scalar model and
those of the fermion model, and show that it can be easily clarified with the
idea of compositeness: the HS field formulation coming from the scalar model is
the `square' of  
the one ensuing from the fermion model, much as the metric is the square of the
vielbein. Finally we analyse the possible obstructions that one may meet
 in constructing the effective action: these are the analogs of anomalies in
ordinary gauge theories. The frame-like formalism allows us to provide explicit
and compact formulas of the latter.

{The effective action method is based on the early  work of Sakharov, 
\cite{Sakharov},
revisited recently by \cite{BCLPS,BCDGLS,BCDGS}. The approach we use is 
based on the worldline quantization method, see 
\cite{Strassler,Segal,Schmidt,Dai, Bonezzi}
and others. As mentioned above here we follow in particular the line of 
\cite{Bekaert:2009ud,Bekaert:2010ky} and \cite{BCDGPS}. }

{The results in this paper are general, they are obtained
abstracting from the idea 
of a locality: an effective field action is mostly nonlocal.} 
In the follow-up of the present paper, denoted II,  \cite{BCDGSII}, we will grow 
bolder. We will abandon
small scale cabotage around the idea of effective action and launch ourselves
in a new enterprise: instead of  trying to derive explicit actions by
integrating out (scalar or fermion) matter fields, we will use the wisdom
(formulas and constructs) acquired in this and previous papers to integrate
$L_\infty$, that is to determine classical (perturbatively) local theories which
automatically satisfy the $L_\infty$ relations and in particular enjoy the HS
gauge invariance/covariance. In II we will exhibit explicit examples of HS
Chern-Simons and HS Yang-Mills theories, both Abelian and non-Abelian. 
As HS YM-like theories contain spin-2 field, they {may} pretend to be 
descriptions of gravity. In the third paper (III) we shall develop {a} HS
geometry-like 
formalism, which is more suitable for a background independent {framework}, and
use it to analyze the spin-2 sector.

\vskip 2cm

{\Large\bf Part I. Functional Equations and Methods}
\vskip 1cm

In this first part of the paper we resume the analysis started in  \cite{BCDGPS}
and develop it in various directions. We first summarize facts about the
fermionic matter model and the relevant effective action. We then complete the
analysis of the $L_\infty$ structure of the latter by considering the case of
the curved $L_\infty$ algebra. We then proceed to some sample calculations of 1-
and 2-point correlator and then show the connection of the perturbative method
outlined in  \cite{BCDGPS} with the more traditional perturbative approach based
on Feynman diagrams.

\section{The method of effective action}

The fermionic matter model is
\be 
S_{matter}&=& \int d^dx \, \overline \psi(i \gamma\!\cdot\! \partial-m)\psi
+\sum_{s=1}^\infty \int  {d^d x}\, J^{(s)}_{a\mu_1\ldots\mu_{s-1}}(x)\,
h_{(s)}^{a\mu_1\ldots\mu_{s-1}}(x)\label{S}\\
&=& S_0 + S_{int}\0
\ee
The interaction part $S_{int}$ can be written in various ways
\be
S_{int}= \langle\!\langle J_a, h^a\rangle\!\rangle = \int   d^dx\,
\frac {d^du}{(2\pi)^d}\, J_a (x,u) h^a(x,u)\label{Sint}
\ee
The (external) gauge fields are collectively represented by  
\be
h^a(x,{u})=\sum_{s=1}^\infty \frac 1{(s-1)!}\, 
h_{(s)}^{a\mu_1\ldots\mu_{s-1} }(x)\, u_{\mu_1}\ldots u_{\mu_{s-1}},\label{hmmm}
\ee 
and
\be
J_a(x,u)&=& \frac {\delta S_{int}}{\delta h^a(x,u)}= \int d^dz \, e^{iz\cdot
u} \overline \psi\left(x+\frac z2\right)  \gamma_a\psi\left(x-\frac
z2\right)\label{Jmu}\\
&=& \sum_{n,m=0}^\infty \frac {(-i)^n i^m}{2^{n+m} n! m!} \,\, \partial^n
\overline \psi(x) \gamma_a \partial^m \psi(x) \,\,\frac
{\partial^{n+m}}{\partial u^{n+m}} \delta(u)\0\\
&=& \sum_{s=1}^\infty{(-1)^{s-1}} J^{(s)}_{a\mu_1 \ldots
\mu_{s-1}}(x)\,\, \frac
{\partial^{s-1}}{\partial u_{\mu_1} \ldots \partial u_{\mu_{s-1}}}\delta(u)\0
\ee
which is obtained by expanding $ e^{iu\cdot z}$. In order to extract
$J^{(s)}_{a\nu_1 \ldots u_{s-1}}(x)$ from $J_a(x,u)$ one must multiply it by
$u_{\nu_1}\ldots u_{\nu_{s-1}}$, integrate over $u$ and divide by $(s-1)!$.
Also
\be
J^{(s)}_{a\mu_1\ldots\mu_{s-1}}(x) = \frac {i^{s-1}} {(s-1)!} \frac
{\partial}{\partial z^{(\mu_1}
}\ldots  \frac {\partial}{\partial z^{\mu_{s-1})}} \overline \psi \left(x+\frac
z2\right)  \gamma_a \psi\left(x-\frac z2\right)
\Big{\vert}_{z=0}.\label{jmmm}
\ee

The gauge transformation of $h_a$ is\footnote{We use the notation $h_a$ instead
of $h_\mu$ as in \cite{BCDGPS} to anticipate and stress the frame-like
interpretation of the master field. See further on.}
\be
\delta_\varepsilon h_a(x,u) = \partial^x_a 
\varepsilon(x,u)-i [h_a(x,u) \stackrel{\ast}{,} \varepsilon(x,u)] 
\equiv {\cal D}^{\ast x}_a \varepsilon(x,u),
\label{deltahxp}
\ee
where we have introduced the covariant derivative
\be
{\cal D}^{\ast x}_a = \partial^x_a - i  [h_a(x,u) \stackrel{\ast}{,}\quad]
.\0
\ee 

The effective action is denoted $\EW[h]$ and takes the form
\be
\EW[h] &=&{\EW[0]+} \sum_{n=1}^\infty\, \frac 1{n!}\int \prod_{i=1}^n d^dx_i\,
\frac {d^du_i}{(2\pi)^d}\,  \EW_{a_1\ldots a_n}^{(n)}(x_1,u_1,\ldots, x_n,
 u_n)\, h^{a_1}(x_1,u_1) \ldots  h^{a_n}(x_n,u_n)\label{EW}\\ 
 &=&{\EW[0]+\sum_{n=1}^\infty\, \frac {i^{n-1}}{n!}\int \prod_{i=1}^n d^dx_i\,
\frac {d^du_i}{(2\pi)^d}\, \langle J_{a_1}(x_1,u_1) \ldots  
J_{a_n}(x_n,u_n)\rangle\, h^{a_1}(x_1,u_1) \ldots  h^{a_n}(x_n,u_n)}\0
\ee 
where
\be
&&\EW_{a_1\ldots a_n}^{(n)}(x_1,u_1,\ldots, x_n, u_n)= 
{i^{n-1}}\langle J_{a_1}(x_1,u_1) \ldots  
J_{a_n}(x_n,u_n)\rangle\label{corrJ1-Jn}\\
&=& \left(\sum^\infty_{s_1=1} \frac {\partial}{\partial u_{1
a^{{(1)}}_1}}\ldots 
\frac {\partial}{\partial u_{1 a^{{(1)}}_{s_1-1}}
}\delta(u_1)\right) \ldots
\left(\sum^\infty_{s_n=1} \frac {\partial}{\partial u_{n
a^{{(n)}}_1}}\ldots 
\frac {\partial}{\partial u_{n
a^{{(n)}}_{s_{{n}}-1}}} \delta(u_n)\right)\0\\
&&\quad\quad \times \, \langle J^{(s_1)}_{a_1 a^{(1)}_{1} \ldots
a^{(1)}_{s_1-1} }\ldots
J^{(s_n)}_{a_n a^{(n)}_{1}(x_1) \ldots a^{(n)}_{s_n-1}}(x_n)\rangle \0
\ee
In the sequel we will assume $\EW[0]=0$, see Appendix \ref{ss:0pt} for a
justification.

The statement of invariance under \eqref{deltahxp} is the global Ward identity
(WI)
\be
\delta_\varepsilon \EW[h]=0\label{WI}
\ee
Taking the variation with respect to $\varepsilon(x,u)$ this becomes 
\be
 \sum_{n=1}^\infty\, \frac 1{n!}\int \prod_{i=1}^n d^dx_i\, \frac
{d^du_i}{(2\pi)^d}\, {\cal D}_x^{\ast a} \EW^{(n+1)}_{a a_1\ldots,
a_n}\!(x,u,x_1,u_1\ldots, x_n, 
u_n)\, h^{a_1}(x_1,u_1) \ldots h^{a_n}(x_n,u_n)=0\label{EWI}
\ee
This must be true order by order in $h$, i.e.
 \be
 0&=&\int \prod_{i=1}^n d^dx_i\, \frac
{d^du_i}{(2\pi)^d}\, {\partial}_x^a \EW^{(n+1)}_{a a_1\ldots
a_n}\!(x,u,x_1,u_1\ldots, x_n, 
u_n)\, h^{a_1}(x_1,u_1) \ldots h^{a_n}(x_n,u_n)\label{EWWI}\\
&&-i\, n \int {\prod_{i=1}^{n-1}} d^dx_i\, \frac
{d^du_i}{(2\pi)^d}\, \Big{[}h^{a}(x,u)\stackrel{\ast}{,}
\EW^{(n)}_{a a_1\ldots a_{n-1}}\!(x,u,x_1,u_1\ldots, x_{n-1}, 
u_{n-1})\Big{]}\0\\
&&\quad\quad\quad\times \, h^{a_1}(x_1,u_1) \ldots
h^{a_{n-1}}(x_{n-1},u_{n-1})\0
\ee
Assuming $\EW^{(1)}_\mu=0$, by functionally differentiating wrt $h^b(y,v)$ and
setting $h=0$,
for $n=1$ this becomes
\be
0=\partial_x^a \EW^{(2)}_{ab}(x,u;y,v).\label{WI1}
\ee
Functionally differentiating also w.r.t. $h^c(z,t)$ (and setting $h=0$) we
get
{
\be
0&=&\frac 1{(2\pi)^d}\left( \partial_z^c \EW^{(3)}_{abc}(x,u;y,v;z,t) 
+ \partial_z^c \EW^{(3)}_{bac}(y,v;x,u;z,t)\right)\0\\
&&-2i\Big{[} \delta(z-x)\delta(t-u)\stackrel{\ast}{,}
\EW^{(2)}_{ab}(z,t;y,v)\Big{]}
-2i \Big{[} \delta(z-y)\delta(t-v)\stackrel{\ast}{,}
\EW^{(2)}_{ba}(z,t;x,u)\Big{]}.\label{WI2}
\ee
}
and so on. In \eqref{WI2}
$\EW^{(2)}_{ab}(x,u;y,v)=\EW^{(2)}_{ba}(y,v;x,u)$, 
due to the cyclic symmetry.

Eq.\eqref{WI} guarantees the gauge invariance of the effective action. 
 The problem now is how to compute the amplitudes $\EW^{(n)}$.
{Here we will not follow the method introduced in
\cite{BCDGPS}, but
mostly the more familiar Feynman diagram method,}
therefore it is {more convenient} to pass to the Fourier
transforms. To start with we
recall that, due to translational invariance,
\be
&&\EW_{a_1\ldots a_n}^{(n)}(x_1,u_1,\ldots, x_n, u_n)= \EW_{a_1\ldots,
a_n}^{(n)}(x_1-x_n,u_1,\ldots, 0, u_n)\0\\
&\equiv& \EW_{a_1\ldots a_n}^{(n)}(x_1-x_n\ldots, x_{n-1}-x_n;u_1,\ldots,
u_n)\0\\
&=&{i^{n-1}}\langle J_{a_1}(x_1,u_1) \ldots  
J_{a_n}(x_n,u_n)\rangle={i^{n-1}}\langle
J_{a_1}(x_1-x_n,u_1) \ldots   J_{a_n}(0,u_n)\rangle\label{transl}
\ee
The Fourier transform of this amplitude is
\be
&&\int d^dx_1 \, e^{ik_1\cdot x_1}\ldots \int d^dx_{n-1}e^{ik_{n-1}\cdot
x_{n-1}}\int d^dx_n e^{-iq\cdot x_n}
\,\EW_{a_1\ldots a_n}^{(n)}(x_1,u_1,\ldots, x_n, u_n)\label{FTWn}\\
&=& \int d^dx_1 \, e^{ik_1\cdot x_1}\ldots  \int d^dx_{n-1}e^{ik_{n-1}\cdot
x_{n-1}} 
\int d^dx_n e^{-iq\cdot x_n}\, \0\\
&& \quad\quad\quad\quad \quad\quad\quad\quad\times \EW_{a_1\ldots
a_n}^{(n)}(x_1\!-\!x_n,\ldots, x_{n-1}\!-\!x_n;u_1,\ldots, u_n)\0\\
&=&\delta\left(q-\sum_{i=1}^{n-1}k_i\right) \int d^dx'_1 \, e^{ik_1\cdot
x'_1}\ldots, \int d^dx'_{n-1} e^{ik_{n-1}\cdot x'_{n-1}} \,\EW_{a_1\ldots
a_n}^{(n)}(x'_1\ldots, x'_{n-1};u_1,\ldots, u_n)\0\\
&=&\delta\left(q-\sum_{i=1}^{n-1}k_i\right) \widetilde \EW_{a_1\ldots
a_n}^{(n)}(k_1\ldots, k_{n-1};u_1,\ldots, u_n)\0\\
&=&{i^{n-1}} \delta\left(q-\sum_{i=1}^{n-1}k_i\right) \bigl{\langle} \widetilde
J_{a_1}(k_1,u_1)\ldots \widetilde J_{a_{n-1}}(k_{n-1},u_{n-1}) \widetilde
J_{a_n}(-q,u_n)\bigr{\rangle}\0
\ee
In particular, for later reference,
\be
\int d^dx \, e^{ik\cdot x} \int d^dy \, {e^{-iq\cdot
y}}\,\EW^{(2)}_{ab}(x,u;y,v)= {\delta(q-k)}
\widetilde\EW^{(2)}_{ab}(k;u,v)={i}{\delta(q-k)} \langle
\widetilde J_a(k,u)
\widetilde J_b(-k,v)\rangle \label{FTW2}
\ee
and
\be
&&\!\!\!\!\!\!\int d^dx \, e^{ik_1\cdot x}\! \int d^dy\, e^{ik_2\cdot y}\!\int
d^dz
\, e^{-iq\cdot z}\,\EW^{(3)}_{abc}(x,u;y,v;z,t)= \delta(q-k_1-k_2)
\widetilde\EW^{(3)}_{abc}(k_1,k_2;u,v,t)\0\\
&&={i^2}\delta(q-k_1-k_2) \big\langle \widetilde J_a(k_1,u) \widetilde
J_b(k_2,v)\widetilde J_c(-q,t)\big\rangle \label{FTW3}
\ee
The WI \eqref{WI1} gets transformed into
\be
k^a\widetilde \EW^{(2)}_{ab}(k;u,v)=i k^a\big\langle \widetilde J_a(k,u)
\widetilde J_b(-k,v)\big\rangle=0 \label{WI1FT}
\ee
while \eqref{WI2} becomes (dropping $i\delta(q-k_1-k_2)$):
{
\be
0&=&i q^c \big\langle \widetilde J_a(k_1,u) \widetilde
J_b(k_2,v)\widetilde J_c(-q,t)\big\rangle + 
i q^c \big\langle \widetilde J_b(k_2,v) \widetilde
J_a(k_1,u)\widetilde J_c(-q,t)\big\rangle \label{WI2FT}\\
&&- 2\Big[\delta\left(t-v+\frac {k_1}2\right) 
\big\langle \widetilde J_a(k_1,u) \widetilde J_b\left(-k_1,t-\frac {k_2}2\right)
\big\rangle\0\\
&&- \delta\left(t-v-\frac {k_1}2\right) 
\big\langle \widetilde J_a(k_1,u) \widetilde J_b\left(-k_1,t+\frac {k_2}2\right)
\big\rangle\0\\
&&+\delta\left(t-u+\frac {k_2}2\right) 
\big\langle \widetilde J_b(k_2,v) \widetilde J_a\left(-k_2,t-\frac {k_1}2\right)
\big\rangle\0\\
&&- \delta\left(t-u-\frac {k_2}2\right) 
\big\langle \widetilde J_b(k_2,v) \widetilde J_a\left(-k_2,t+\frac {k_1}2\right)
\big\rangle\Big]\0
\ee
}
Using \eqref{FTW2} and \eqref{FTW3} one can write  this as
{
\be
0&=&q^c\left(\widetilde \EW^{(3)}_{abc}(k_1,k_2;u,v,t) 
+\widetilde \EW^{(3)}_{bac}(k_2,k_1;v,u,t)\right)
\label{WI2FT'}\\
&&-2\Big[ \delta\left(t-v+\frac {k_1}2\right)\widetilde \EW^{(2)}_{ab}
\left(k_1;u,t-\frac {k_2}2\right)
-\delta\left(t-v-\frac {k_1}2\right)\widetilde \EW^{(2)}_{ab}
\left(k_1;u,t+\frac {k_2}2\right)\0\\
&& +\delta\left(t-u+\frac {k_2}2\right)\widetilde \EW^{(2)}_{ba}
\left(k_2;v,t-\frac {k_1}2\right)
-\delta\left(t-u-\frac {k_2}2\right)\widetilde \EW^{(2)}_{ba}
\left(k_2;,v,t+\frac {k_1}2\right)\Big]\0
\ee
}
To extract the WI in components one must multiply by a suitable expression 
$\frac {\partial ^j}{\partial u^j} \delta(u)$ and integrate over  $u$. Similarly
for $v$ and $t$.

For a derivation of \eqref{WI2FT} and \eqref{WI2FT'}, see Appendix
\ref{s:WI2FT}{.}

\section{The case  $\EW^{(1)}_a\neq 0$}

\subsection{A curved $L_\infty$ algebra}

It was shown in \cite{BCDGPS} that, underlying  the effective action obtained by
integrating out a fermion field coupled to external sources, there is an
algebraic structure, which is unveiled once we consider the relevant equations
of motion. The basic relations in this game are the eom's
 \be
\EF_a(x,u)=0\label{GenEoM}
\ee
where
\be \label{gEoMt}
\EF_a(x,u) &\equiv&  \sum_{n=0}^\infty\, \frac 1{n!}\int \prod_{i=1}^{n}
d^dx_i\,
\frac
{d^du_i}{(2\pi)^d}\,  \EW_{a a_1\ldots a_{n}}^{(n+1)}(x,u,x_1,u_1,\ldots,
x_{n},  u_{n}) \0\\
&&\quad \times \, h^{a_1}(x_1,u_1) \ldots  h^{a_{n}}(x_{n},u_{n})
\0
\ee 
 and the covariance relation
\be
\delta_\varepsilon \EF_a (x,u)=i [ \varepsilon(x,u) \stackrel{\ast}{,} 
\EF_a(x,u)]\label{dvareEF}
\ee
which can also be rewritten as follows
\be
i [\varepsilon \stackrel{\ast}{,}\langle\!\langle \EW_{a}^{(n+1)}\, ,\,
h^{\otimes n}\rangle\!\rangle]&=&\frac 1{n+1}\sum_{i=1}^{n+1} \langle\!\langle
\EW_{a a_1\ldots
\mu_i\ldots a_{n+1}}^{(n+2)}\,
,\,h^{a_1}\ldots\partial_x^{a_i}\varepsilon\,  \ldots h^{a_{n+1} }
\rangle\!\rangle\label{EFequations}\\
&& - i \, \sum_{i=1}^{n} 
\langle\!\langle \EW_{a a_1\ldots a_i\ldots \mu_n}^{(n+1)}\, ,\,
h^{a_1}\ldots\,[h^{a_i}\stackrel{\ast}{,}
\varepsilon]\ldots h^{a_n}\rangle\!\rangle \0
\ee

It was shown in section 3 of  \cite{BCDGPS}, that this allows us to define
multilinear maps (products) 
$L_j$, $j=1,\ldots,\infty$ of degree $d_j=j-2$ among vector spaces $X_i$ of
degree $i$, defined by the assignments  $\varepsilon\in X_0$, $h_a\in X_{-1}$
and  ${\EF}_a \in X_{-2}$, which satisfy  the relations
\be
\sum_{i+j=n+1} (-1)^{i(j-1)} \sum_\sigma (-1)^\sigma \epsilon(\sigma;x)\, L_j
(L_i(x_{\sigma(1)},\ldots, x_{\sigma(i)}),x_{\sigma(i+1)},\ldots
,x_{\sigma(n)})=0
\label{Ln}
\ee
In this formula $\sigma$ denotes a permutation of the entries so that
$\sigma(1)<\ldots\sigma(i)$ and $\sigma(i+1)<\ldots\sigma(n)$, and
$\epsilon(\sigma;x)$ is the Koszul sign. 
Nonvanishing examples of such products are
\be
L_1 (\varepsilon)^a&=& \partial_x^a \varepsilon(x,u)\label{ell1e}\\
L_2 (\varepsilon,h)^a&=& -i [h^a(x,u)\stackrel{\ast}{,}
\varepsilon(x,u)]=- L_2 (h,\varepsilon)^a\label{ell2eh}\\
L_2(\varepsilon_1,\varepsilon_2)&=&  i\, [{\varepsilon_1}\stackrel{\ast}{,}
{\varepsilon_2}] \label{l2e1e2}
\ee
together with
\be
L_1(h) = \langle\!\langle \EW_a^{(2)} ,h\rangle\!\rangle = 
\int  d^dx_i\, \frac
{d^du_i}{(2\pi)^d}\,  \EW_{a a_1}^{(2)}(x,u,x_1,p_1)
 h^{a_1}(x_1,u_1)\label{ell1h}
\ee
and $L_n(h_1,\ldots,h_n)$, see the definition in \cite{BCDGPS}. Assuming the
one-point correlator $\EW^{(1)}_a= 0$, in \cite{BCDGPS} it was proved that the
relation \eqref{Ln} are satisfied.

In many cases, however $\EW^{(1)}_a\neq 0$. When this happens we have to
introduce
an additional `product' $L_0$, 
besides the $L_n$ of \cite{BCDGPS}, \cite{Stasheff1}. The algebra in this case
is called {\it
curved} $L_\infty$\footnote{{The notion of curved $L_\infty$
algebra has been suggested to us by Jim Stasheff.}}. 
$L_0$ is defined simply by setting
\be
L_0= \EW^{(1)}_a\label{L0}
\ee
Both sides of this equation have degree -2, coherently with the fact that the
degree of $L_n$ is $n-2$.
In this case  $L_1$ is not nilpotent and the defining property 
$L_1^2=0$ of the $L_\infty$ algebra is modified as follows
\be
L_1 \left( L_1(v)\right) + L_2\left( L_0,v\right)=0 \label{L1L1+L2L0}
\ee 
where $v\in X= X_0\oplus X_{-1}\oplus X_{-2}$. Due to the degree counting, this
relation 
is only nontrivial when $v\in X_0$, i.e. when $v$ is $\varepsilon$. Now using
eq.(3.21) of 
\cite{BCDGPS}, i.e.
\be
L_2(\varepsilon,E) = i [\varepsilon \stackrel{\ast}{,} E]
\ee
where $E$ represents $\EF_a$ or any of its homogeneous  {pieces}, and
 recalling that $L_1(\varepsilon)^a(x,u)= \partial_x^a
\varepsilon(x,u)$ and 
$L_1(h)_a=\langle\!\langle \EW_a^{(2)}, h\rangle\!\rangle$, this equation
becomes
\be
i [ \EW^{(1)}_a,\varepsilon] + \langle\!\langle \EW_{ab} ^{(2)}
h^b\rangle\!\rangle =0
\label{Linfcurv}
 \ee
This is precisely the case $n=0$ of \eqref{EFequations}. All the other
$L_\infty$ defining 
relations remain unchanged, because, for instance, the relation
\be
L_3L_0 - L_2L_1+L_1L_2=0\label{L3L0}
\ee
is not a priori excluded by the degree counting, but in \cite{BCDGPS} we have 
proved that $L_3(E,*,*)=0$ is consistent for 
$E$ of degree -2. And so on. 

$L_0$ is called the {\it curvature} of the curved $L_\infty$ algebra.

In this case \eqref{WI1FT} in momentum space becomes
\be
k^a\widetilde{\EW}^{(2)}_{ab}(k;u,v)+ \delta(u-v)
\left(\EW^{(1)}_b \left(0, u+\frac k2\right) - \EW^{(1)}_b \left(0, u-\frac
k2\right)\right)=0
\label{WI1FT2}
\ee
In this formula $\EW^{(1)}_b (0, u)\equiv\EW^{(1)}_b (x=0, u)$.

 \subsection{WI for 2-pt functions}

{The method introduced in \cite{BCDGPS} to calculate the current
amplitudes is less manageable then the more familiar method based on Feynman
diagrams. It is however more compact and may turn out to be useful in some
instances. Here we would like to show in a simple example, the WI for 2-pt
correlators, that it leads to the expected results. }

For $n=1$  and assuming $\EW^{(1)}_a\neq 0$, (\ref{WI1}) becomes
\be
0=\frac{1}{(2\pi)^d}\partial_x^a \EW^{(2)}_{ab}(x,u;y,u')-i \Big{[}
\delta(x-y)\delta(u-u')\stackrel{\ast}{,}
\EW^{(1)}_{b}(x,u)\Big{]}\label{EWWI1}
\ee
where
\be
\EW_{a_1,\ldots, a_n}^{(n)}(x_1,u_1,\!\!&\ldots&\!\!, x_n, 
u_n)= -N{ \frac {n!}n } \int_\epsilon ^\infty dt \, \int
\prod_{i=1}^n
d^dq_i \int_{-\infty}^{\infty} \frac{d\omega}{2\pi }\, e^{i\omega t}
\0\\
&\times&\! {\tr} \Big{[} \gamma_{a_1} \frac 1{\slashed{ q}_1+m'}
\gamma_{a_2} \ldots \gamma_{a_{n-1}}\frac 1{ \slashed{q}_{n-1}+m'}
\gamma_{a_n} \frac 1{ \slashed{ q}_{n}+m'}\Big{]}  \0\\
&\times&\! \prod_{j=1}^n e^{i q_j\cdot {\left(x_j 
-x_{j+1}\right) } }\,\,
\delta\!\left(u_1-\frac {q_1+q_n}{2}\right)\ldots   \delta\!\left(u_n-\frac
{q_{n-1}+q_n}2\right)\label{EWn}
\ee
and we introduced $m'= m- i( \omega- i \varepsilon)$.
In particular, for $n=1$ (\ref{EWn}) becomes
\be
\EW_{b}^{(1)}(x,u)&=& -N \int_\epsilon ^\infty dt \, \int
d^dq \int_{-\infty}^{\infty} \frac{d\omega}{2\pi }\, e^{i\omega t}
\, {\tr} \Big{[} \gamma_{b} \frac 1{\slashed{ q}+m'}\Big{]}  
\delta\!\left(u-q\right)\0\\
&=& -2^{\lfloor \frac d2\rfloor}\,i\, N \int_{-\infty}^{\infty}
\frac{d\omega}{2\pi }\, \frac{e^{i\omega
\epsilon}}{\omega}\frac{u_b}{p^2-m'^2}\label{EW1}
\ee
Note that the above equation does not depend on $x$. We can write
$\EW_{\nu}^{(1)}(x,u)=\EW_{\nu}^{(1)}(u)$.
Forr $n=2$
\be
\EW_{ab}^{(2)}(x,u,y,u')&=& -N \int_\epsilon ^\infty dt \, \int
d^dq_1 d^dq_2\int_{-\infty}^{\infty} \frac{d\omega}{2\pi }\, e^{i\omega t}
 {\tr} \Big{[} \gamma_{a} \frac 1{\slashed{ q}_1+m'}
\gamma_{b}\frac 1{ \slashed{q}_{2}+m'}\Big{]}  \0\\
&\times&\! e^{i (q_1-q_2)\cdot {\left(x 
-y\right) } }\,\,
\delta\!\left(u-\frac {q_1+q_2}{2}\right)   \delta\!\left(u'-\frac
{q_1+q_2}2\right)
\ee
We can introduce new variables $q_1-q_2=k$ and $q_1=q$. Then, (\ref{EW2}) reads
\be
\EW_{ab}^{(2)}(x,u,y,u')&=& -N \int_\epsilon ^\infty dt \, \int
d^dq d^dk\int_{-\infty}^{\infty} \frac{d\omega}{2\pi }\, e^{i\omega t}
 {\tr} \Big{[} \gamma_{a} \frac 1{\slashed{ q}+m'}
\gamma_{b}\frac 1{ \slashed{q}-\slashed{k}+m'}\Big{]}  \0\\
&\times&\! e^{i k\cdot {\left(x 
-y\right) } }\,\,
\delta\!\left(u-\frac {2q-k}{2}\right)   \delta\!\left(u'-\frac
{2q-k}2\right)
\ee
After the evaluation of the trace and the integration over $t$ and $q$ we
obtain 
\be
\EW_{ab}^{(2)}(x,u,y,u')&=& -2^{\lfloor \frac d2\rfloor}\,i\, N   \int
d^dk\int_{-\infty}^{\infty} \frac{d\omega}{2\pi }\, \frac{e^{i\omega
\epsilon}}{\omega}
\frac{1}{\left(\left(u+\frac k2\right)^2-m'^2\right)\left(\left(u-\frac
k2\right)^2-m'^2\right)}\0\\
&\times&\! \left[2u_a u_b-\frac{k_a
k_b}{2}-\left(u^2-\frac{k^2}{4}-m'^2\right)\eta_{ab}\right]e^{i k\cdot
{\left(x 
-y\right) } }\,\,
\delta\!\left(u-u'\right)   \label{EW2}
\ee

Let us now focus on the first term in (\ref{EWWI1})

\be
\frac{1}{(2\pi)^d}\partial_x^a \EW^{(2)}_{ab}(x,u;y,u')&=& -2^{\lfloor
\frac d2\rfloor}\,i\, N   \int
\frac{d^dk}{(2\pi)^d}\int_{-\infty}^{\infty} \frac{d\omega}{2\pi }\,
\frac{e^{i\omega \epsilon}}{\omega}
\frac{1}{\left(\left(u+\frac k2\right)^2-m'^2\right)\left(\left(u-\frac
k2\right)^2-m'^2\right)}\0\\
&\times&\! \left[2u_a u_b-\frac{k_a
k_b}{2}-\left(u^2-\frac{k^2}{4}-m'^2\right)\eta_{ab}\right]\partial_x^a
e^{i k\cdot {\left(x 
-y\right) } }\,\,
\delta\!\left(u-u'\right) \0\\
&=&  2^{\lfloor \frac d2\rfloor}\, N   \int
\frac{d^dk}{(2\pi)^d}\int_{-\infty}^{\infty} \frac{d\omega}{2\pi }\,
\frac{e^{i\omega \epsilon}}{\omega} e^{i k\cdot {\left(x-y\right) } }\,\,
\delta\!\left(u-u'\right)\0\\
&\times&\!  \frac{\left[2(u\cdot k)\,
p_b+\left(m'^2-u^2-\frac{k^2}{4}\right)k_b\right]}{\left(\left(u+\frac
k2\right)^2-m'^2\right)\left(\left(u-\frac k2\right)^2-m'^2\right)}
\label{EWWI1first}
\ee

On the other hand, the second term in (\ref{EWWI1}) is
\be 
-i \Big{[} \delta(x-y)\delta(u-u')\stackrel{\ast}{,}
\EW^{(1)}_{b}(x,u)\Big{]}&=&\delta(x-y)\delta(u-u')\sum_{n=0}^{\infty}\frac{1}
{(2n+1)!}\left(\frac{i}{2}\right)^{2n}\left(\stackrel{\leftarrow}{\partial}
_x\cdot\stackrel{\rightarrow}{\partial}_u\right)^{2n+1}\EW^{(1)}_{b}(u)\0\\
&=& 2^{\lfloor \frac d2\rfloor}\, N   \int
\frac{d^dk}{(2\pi)^d}\int_{-\infty}^{\infty} \frac{d\omega}{2\pi }\,
\frac{e^{i\omega \epsilon}}{\omega}e^{i k\cdot
{\left(x-y\right)}}\delta\!\left(u-u'\right)\0\\
&\times&\!
\sum_{n=0}^{\infty}\frac{1}{2^{2n}(2n+1)!}\left(k\cdot\stackrel{\rightarrow}{
\partial}_u\right)^{2n+1}\left(\frac{u_b}{u^2-m'^2}\right)
\ee
Taking the derivatives w.r.t. $u$ gives
\be 
-i \Big{[} \delta(x-y)\delta(u-u')\stackrel{\ast}{,}
\EW^{(1)}_{b}(x,u)\Big{]}&=& 2^{\lfloor \frac d2\rfloor}\, N   \int
\frac{d^dk}{(2\pi)^d}\int_{-\infty}^{\infty} \frac{d\omega}{2\pi }\,
\frac{e^{i\omega \epsilon}}{\omega}e^{i k\cdot
{\left(x-y\right)}}\delta\!\left(u-u'\right)\0\\
&\times& \sum_{n=0}^{\infty}\sum_{j=0}^{n}\left(-4^{j-n}(k\cdot
u)^{2j}k^{2(n-j)}(m'^2-u^2)^{-2-n-j}\right)\0\\
&\times&\left(\binom{n+j}{2j}(m'^2-u^2)k_b+2\binom{n+j+1}{2j+1}(k\cdot
u)u_b\right)
\ee
This can be resummed into 
\be 
-i \Big{[} \delta(x-y)\delta(u-u')\stackrel{\ast}{,}
\EW^{(1)}_{b}(x,u)\Big{]}&=&- 2^{\lfloor \frac d2\rfloor}\, N   \int
\frac{d^dk}{(2\pi)^d}\int_{-\infty}^{\infty} \frac{d\omega}{2\pi }\,
\frac{e^{i\omega \epsilon}}{\omega} e^{i k\cdot {\left(x-y\right) } }\,\,
\delta\!\left(u-u'\right)\0\\
&\times&\!  \frac{\left[2(u\cdot k)\,
p_b+\left(m'^2-u^2-\frac{k^2}{4}\right)k_b\right]}{\left(\left(u+\frac
k2\right)^2-m'^2\right)\left(\left(u-\frac k2\right)^2-m'^2\right)} 
\ee
Now, this term has exactly the opposite sign w.r.t. (\ref{EWWI1first}) and we
see
that (\ref{EWWI1}) is indeed valid for 2-pt functions.

\section{Explicit Calculations}

After studying the general properties of the current amplitudes in the previous
sections, the next step involves the methods to explicitly compute such
amplitudes. In fact in \cite{BCDGPS} rather explicit formulas were presented.
Using such formulas in Appendix \ref{s:0and1pt} we have calculated the 0- and
1-pt current correlators. However, to compute more complicated correlators we
find it more {practical} to resort to the usual Feynman rules.

\subsection{Feynman rules}
\label{s:Feynmanrules}

The free part of the action \eqref{S} gives rise to the usual fermion propagator
\be
\frac i{\slashed{p}+m} \label{fermionprop}
\ee
The interaction part $S_{int}$ is given by \eqref{Sint}, $J_a(x,u)$ is
given by \eqref{Jmu}
and $J^{(s)}_{a\mu_1 \ldots \mu_{s-1}}(x)$ is defined by \eqref{jmmm}.
Therefore
\be
S_{int} &=&\int d^dx \, \sum_{n,m=0}^\infty \frac {i^n (-i)^m}{2^{n+m} n! m!}\,
\partial^n \overline \psi(x) \gamma_a \partial^m \psi(x) 
h^{a\mu_1\ldots\mu_n\mu_{n+1}\ldots\mu_{n+m}}(x)\label{Sint2}\\
&=& \int d^dx \, \sum_{s=1}^\infty \sum_{n=0}^{s-1}\frac {(-1)^n
(-i)^{s-1}}{2^{s-1} n! (s-n-1)!} \,\partial_{\mu_1}\ldots \partial_{\mu_n}
\overline \psi(x) \gamma_a \partial_{\mu_{n+1}}\ldots \partial_{\mu_{s-1}}
\psi(x) \,\,
h^{a\mu_1\ldots\mu_{s-1}}(x)\0
\ee
By replacing the fields with plane waves while keeping the tensor structure we
can determine the $V_{ffh}$ vertex:
\be
V_{ffh}: \quad \frac { i}{2^{s-1}{(s-1)!}},\gamma_\mu \, 
\sum_{n=0}^{s-1}\,k'_{(\mu_1}\ldots
k'_{\mu_n}  k_{\mu_{n+1}}\ldots k_{\mu_{s-1})}\label{Vffh}
\ee
where $k$ is an incoming fermion momentum, while $k'$ is an outgoing one. The
momentum conservation ($\delta(q+k-k')$ , $q$ incoming) at the vertex is
understood. The vertex \eqref{Vffh} can be easily derived from \eqref{Sint} and
the first line of \eqref{Jmu}, by replacing $\overline \psi\left(x+\frac
z2\right)$ with
$e^{-ik' \cdot \left(x+\frac z2\right)}$, $\psi\left(x-\frac z2\right)$ with
$e^{ik \cdot \left(x-\frac z2\right)}$ and $h^a(x,u)$ with $h^a(u)
e^{iq\cdot x}$. Performing the integrals one gets
\be
&&i\delta(q+k-k') \int d^dz\,\frac {d^dp}{(2\pi)^d}\, e^{i \left( p- \frac
{k+k'}2\right)\cdot z}
\gamma_a h^a(p)=i\delta(q+k-k')\gamma_a h^a\left( \frac {k+k'}2\right)\0\\
&=& i\delta(q+k-k') \gamma_a \sum_{n=0}^\infty \frac 1{2^n
n!}h^{a\mu_1\ldots \mu_n}
(k+k')_{\mu_1}\ldots (k+k')_{\mu_n}\label{vertex}
\ee
where $h^{a\mu_1\ldots \mu_n}$ is a constant tensor completely symmetric in
$\mu_1,\ldots ,\mu_n$. So one can write the vertex also as
\be
V_{ffh}:\quad V^{(s)}_{a\mu_1 \ldots\mu_{s-1}}(k+k')= \frac {
i}{2^{s-1}{(s-1)!}}
\gamma_{a} (k+k')_{(\mu_1}\ldots (k+k')_{\mu_{s-1})}\label{Vffh'}
\ee
or, introducing two polarization vectors $m^a, n^\nu$,
\be
V_{ffh}:\quad  {V^{(s)}(m,n;k+k')= \frac {
i}{2^{s-1}{((s-1)!)^2}}\,  \slashed{m}\,
\left(n\!\cdot\! (k+k')\right)^{s-1}}\label{Vffh''}
\ee
We can also introduce the global vertex{ 
\be
\EV(m,n;k+k') &=& \sum_{s=1}^\infty  V^{(s)}(m,n;k+k')=  i\sum_{s=1}^\infty\frac
1{2^{s-1} ((s-1)!)^2}\,  \slashed{m}\, \left(n\!\cdot\! (k+k')\right)^{s-1}\0\\
&=&i\,\slashed{m} \,I_0\left(2 n\!\cdot\!(k+k') \right)\label{Vffhglobal}
\ee}
where $I_0(z)$ is the modified Bessel function of second kind.

Next, in order to derive the Feynman rules, we couple the fermions to external
currents $j,\bar j$ via an action term $\int \left(\bar j \psi + \bar \psi
j\right)$ and define
\be
Z[h;\bar j,j] &=& \int {\cal D}\bar \psi {\cal D}\psi \, e^{i\left(S_0 +
\int\left(\bar j \psi + \bar \psi j\right)\right)} 
e^{i S_{int}} = e^{i S_{int} \left(\bar \psi = {i}\frac
{\delta}{\delta j}, \psi= { -i}
\frac {\delta}{\delta \bar j}\right) }\, {e^{ \int \bar j P
j}}\label{Z}
\ee
where $P$ is the propagator. Then we define the generator of connected Green
functions
\be
W[h;j,\bar j] = -i \log Z[h;\bar j,j], \quad\quad W[h]= W[h;j,\bar j]
\Big{\vert}_{j=\bar j=0}\label{Wjbarj}
\ee

Now, the connected Green functions of currents are given by
\be
&&{i^{n-1}}\langle J^{(s_1)}_{\mu_1^{(1)}\ldots \mu_{s_1}^{(1)}}(x_1)\ldots 
J^{(s_n)}_{\mu_1^{(n)}\ldots \mu_{s_n}^{(n)}}(x_n)\rangle =\frac {\delta}{\delta
h^{\mu_1^{(1)}\ldots \mu_{s_1}^{(1)}}(x_1)}\ldots
\frac {\delta}{\delta h^{\mu_1^{(n)}\ldots \mu_{s_n}^{(n)}}(x_n)}
W[h]\Big{\vert}_{h=0}\label{J1Jn}\\
&=& {\frac {i^{n-1}}{n!n!}}\frac {\delta}{\delta
h^{\mu_1^{(1)}\ldots
\mu_{s_1}^{(1)}}(x_1)}\ldots
\frac {\delta}{\delta h^{\mu_1^{(n)}\ldots \mu_{s_n}^{(n)}}(x_n)} \Bigg{(}
S_{int} \left( {i} \frac {\delta}{\delta j},  {-i} \frac
{\delta}{\delta \bar j};
h\right)^n \left(\int \bar j P j \right)^n \Bigg{)}_c\,\Bigg{\vert}_{ h,j,\bar
j=0}\0
\ee
where the subscript $c$ means that only the connected graphs are retained.

The first contribution is the tadpole or one-point function
\be
 { \Big{\langle} n^{s-1} \!\cdot \! m\!\cdot \!J^{(s)}(0) \Big{\rangle}
={-}
\int \frac{d^dp}{(2\pi)^d}\tr \left( \frac
{{1}}{\slashed{p}+m}V^{(s)}(m,n;
2p)\right)} \label{tadpole}
\ee
where $n^s \!\cdot \! m\!\cdot \!\widetilde J^{(s)}(k)$ is the Fourier transform
of
$n^{\nu_1} \ldots n^{\nu_s} m^a J^{(s)}_{a \nu_1\ldots \nu_{s-1}}(x)$

The two-point function is
\be
&&  \Big{\langle} n_1^{s_1} \!\cdot \!m_1\!\cdot\! \widetilde
J^{(s_1)}(k)\,\, n^{s} \!\cdot \! m \!\cdot \!\widetilde
J^{(s)}(-k)\Big{\rangle}\0\\
& =& {- \int\frac{d^dp}{(2\pi)^d}\tr
\left( \frac 1{\slashed{p}+m} V^{(s)}(m_1,n_1; 2p-k)
\frac 1{\slashed{p}-\slashed{k}+m}V^{(s_1)}(m,n; 2p-k)\right)}\label{bubble}
\ee

The three-point function is ($q=k_1+k_2$)
 {
\be
&& \Big{\langle} n_1^{s_1} \!\cdot \!m_1 \!\cdot \!
\widetilde J^{(s_1)}(k_1)\,\, n_2^{s_2} \!\cdot \! m_2 \!\cdot \!\widetilde
J^{(s_2)}(k_2)\,\, n^{s} \!\cdot \! m \!\cdot \!\widetilde
J^{(s)}(q)\Big{\rangle}\label{triangle}\\
& =&\!\!{-} \int\frac{d^dp}{(2\pi)^d}\tr \left( \frac
1{\slashed{p}+m}
V^{(s_1)}(n_1; 2p-k_1)
\frac 1{\slashed{p}-\slashed{k}_1+m}V^{(s_2)}(n_2; 2p-2k_1-k_2)\frac
1{\slashed{p}-\slashed{q}+m}V^{(s)}(n; 2p-q)\right)\0
\ee
}

The divergences of these amplitudes can be evaluated by acting with $k\!\cdot\!
\frac {\partial}{\partial n}$ in the case of \eqref{bubble} and with $q\!\cdot\!
\frac {\partial}{\partial n}$ in the case of \eqref{triangle}.

\subsection{{A master field derivation of Feynman rules}}

It is interesting to observe that one can get Feynman rules not only for the
component currents, but for the master current $J_\mu(x,u)$ itself.  One
proceeds as follows. The free action
can be written
\be
S_0&=& \int d^dx \, \overline \psi(x)(i \gamma\!\cdot\!
\partial-m)\psi(x)\label{S0alt}\\
&=& \int d^d x \, d^dz {\frac{d^du}{(2\pi)^d}}\, e^{iz\cdot u}\,\overline
\psi\left(x+\frac z2\right)(i \gamma\!\cdot\! \partial-m)\psi\left(x-\frac
z2\right)\0
\ee
while the interaction part is
\be
S_{int}&=& \langle\!\langle J_a, h^a\rangle\!\rangle = \int   d^dx\,
\frac {d^du}{(2\pi)^d}\, J_a (x,u) h^a(x,u)\label{Sintb}\\
&=&\int d^d x \, d^dz {\frac{d^du}{(2\pi)^d}}\, e^{iz\cdot u}\, 
\psi\left(x+\frac z2\right) \gamma_a\psi\left(x-\frac
z2\right)\, h^a (x,u)\0
\ee
As before we define the $Z$ function
\be
Z[h;\bar j,j] &=& \int {\cal D}\bar \psi {\cal D}\psi \, e^{i\left(S_0 +
\int\left(\bar j \psi + \bar \psi j\right)\right)} 
e^{i S_{int}},\label{Sintalt}
\ee
where, however, now
\be
\int\left(\bar j \psi + \bar \psi j\right)= \int d^d x \, d^dz
 {\frac{d^du}{(2\pi)^d}}\, e^{iz\cdot u}\, 
\left( \overline j\left(x+\frac z2\right) \psi\left(x-\frac z2\right) +
\overline \psi\left(x+\frac z2\right) j\left(x-\frac z2\right)
\right)\label{Sintalt1}
\ee
The factor $e^{iz\cdot u}$ enters the definition of the integration measure over
$x,u$ and $z$. Next, as usual, one completes the square  and integrates over
$\psi$ and $\overline\psi$. As a result, apart from an overall constant, we get 
\be
 Z[h;\bar j,j] = e^{i S_{int} \left(\bar \psi = {i} \frac
{\delta}{\delta j}, \psi= 
{-i}\frac {\delta}{\delta \bar j}\right) }\,{ e^{ \int \bar j P
j}}\label{Zalt}
\ee
where
\be 
\int \overline j P j= \int d^d x_1 d^dx_2 d^dz{\frac{d^du}{(2\pi)^d}}\,
e^{iz\cdot u}\, \overline j\!\left(x_1-\frac z2\right)\, {  P}\left(x_1-\frac
z2,x_2+\frac z2\right)\,\, j\!\left(x_2+\frac z2\right)\label{jPj}
\ee
and $  P$ is the propagator:
\be
{ P}(x,y) = {i}\int \frac{d^dk}{(2\pi)^d}\frac{e^{ik\cdot(x-y)}}{{\slashed
k}+m}\label{propxy}
\ee
The generator of connected Green functions is defined as above
\be
W[h;j,\bar j] = -i \log Z[h;\bar j,j], \quad\quad W[h]= W[h;j,\bar j]
\Big{\vert}_{j=\bar j=0}\label{Wjbarj1}
\ee
The correlators of the master current $J_a(x,u)$ are obtained by
differentiating $W$ with respect to $h^a(x,u)$. The Feynman rules have the
same combinatorics as in an ordinary field theory, but the propagators and
vertices are different.

\subsubsection{1-point function}
{
\be
\langle J_a(x,u)\rangle &=&  \int d^dz \, e^{iz\cdot u}\langle \overline\psi
\left(x+\frac z2\right)\gamma_a \psi\left(x-\frac z2\right)\rangle 
= \int d^dz \, e^{iz\cdot u} \frac {\delta}{\delta j\left(x+\frac
z2\right)}\gamma_a 
\frac{\delta}{\delta \overline j\left(x-\frac z2\right)}\0\\ 
&&\times\int d^dy_1 d^dy_2 \frac{d^dv}{(2\pi)^d}\, d^dt\, e^{iv\cdot t}
\,\overline j\!\left(y_1-\frac t2\right)\, {  P}\left(y_1-\frac t2,y_2+\frac
t2\right)\,\, j\!\left(y_2+\frac t2\right)\label{1p0}\\
&=&- \int d^dz \, e^{iz\cdot u}\int d^dy_1 d^dy_2 \frac{d^dv}{(2\pi)^d}\,
d^dt\, e^{iv\cdot t} \,\tr \left(\gamma_a {  P}\left(y_1-\frac t2,y_2+\frac
t2\right)\right) \0\\
&& \times\, \delta \left(x-y_1-\frac {z-t}2 \right) \,\delta \left(x-y_2+\frac
{z-t}2 \right) \0
\ee
}
Replacing \eqref{propxy} and simplifying
\be
\langle J_a(x,u)\rangle ={-\int \frac{d^dk}{(2\pi)^d}\, d^dz\, 
e^{i(u-k)\cdot z}\,\tr \left(\gamma_a\frac i{{\slashed k}+m} \right) }
=- \tr \left(\gamma_a \frac {{i}}{{\slashed u}+m}  \right)
=-\,2^{\lfloor\frac
d2\rfloor}{i}  \frac {u_a}{u^2-m^2}\label{1pt1}
\ee
Now, inverting \eqref{Jmu}, one gets the one-point correlators for component
currents
\be
\langle J^{(s)}_{a\mu_1\ldots \mu_{s-1}}(x)\rangle&=&  \frac {
{{1}}}{(s-1)!} \int \frac{d^du}{(2\pi)^d}\, \langle
J_a(x,u)\rangle \, 
u_{\mu_1} \ldots u_{\mu_{s-1}}\0\\
&=&  \frac { {1}}{(s-1)!} \int\frac{d^du}{(2\pi)^d}\,\frac {
{2^{\lfloor\frac d2\rfloor}}u_a}{u^2-m^2} \, u_{\mu_1} \ldots
u_{\mu_{s-1}}\label{1p2}
\ee  
{The Fourier transform of this amplitude corresponds to
(\ref{tadpole}).}
\subsubsection{2-point function}

The connected 2-pt function for master currents is
\be
\langle J_a(x,u)J_b(y,v)\rangle &=&\int d^dz \, e^{iz\cdot u} \frac
{\delta}{\delta j\left(x+\frac z2\right)}\gamma_a \frac{\delta}{\delta
\overline j\left(x-\frac z2\right)}\int d^dt \, e^{it\cdot v} \frac
{\delta}{\delta j\left(y+\frac t2\right)}\gamma_b 
\frac{\delta}{\delta \overline j\left(y-\frac t2\right)}\0\\
&&\times {\frac 14 } \int\frac {d^du_1}{(2\pi )^d} \,  d^dz_1 \,
e^{iz_1\cdot u_1}\int dx_1 dx_2\, \overline j\left(x_1-\frac {z_1}2\right) 
P_{1,2}\,j\left(x_2+\frac {z_1}2\right)\0\\
&&\times \int \frac {d^du_2}{(2\pi )^d} \,  d^dz_2 \, e^{iz_2\cdot u_2}\int dx_3
dx_4\, \overline j\left(x_3-\frac {z_2}2\right)  P_{3,4}\,j\left(x_4+\frac
{z_2}2\right)\label{2pt0}
\ee
where
\be
{ P}_{1,2} = {i}\int \frac{d^dk_1}{(2\pi
i)^d}\frac{e^{ik_1\cdot(x_1-x_2-z_1)}}{{\slashed k_1}+m},\quad\quad { P}_{3,4}
={i}\int \frac{d^dk_2}{(2\pi
i)^d}\frac{e^{ik_2\cdot(x_3-x_4-z_2)}}{{\slashed
k_2}+m}\label{propxy1234}
\ee
After integrating the delta functions coming from the $j, \overline j$
differentiations, one gets
\be
&&\langle J_a(x,u)J_b(y,v)\rangle ={-}\int d^dz \, d^dt\frac {d^du_1}{(2\pi
)^d} \frac {d^du_2}{(2\pi )^d}  d^dz_1 \,  d^dz_2 \,  e^{i(z\cdot u+t\cdot v)}\,
 e^{i(z_1\cdot u_1+ z_2\cdot u_2)}
\label{2pt1}\\
&&\times\int \frac {d^dk_1}{(2\pi )^d} \frac {d^dk_2}{(2\pi
)^d}\, { e^{(k_1-k_2)\cdot(x-y)}e^{-i(k_1+k_2)\frac{t+z}2}}\tr
\left( \gamma_a\frac {{i}}{{\slashed k_1}+m} \gamma_b \frac
{{i}}{{\slashed
k_2}+m}\right)\0\\
&=&{-}\int d^dz \, d^dt\,\int \frac {d^dk_1}{(2\pi )^d} \frac {d^dk_2}{(2\pi
)^d}\,{
e^{(k_1-k_2)\cdot\left(x-y\right)}e^{iz\left(u-\frac{k_1+k_2}2\right)}e^{
it\left(v-\frac{k_1+k_2}2\right)}}\tr \left( \gamma_a\frac
{{i}}{{\slashed k_1}+m}
\gamma_b \frac {{i}}{{\slashed k_2}+m}\right)\0
\ee
setting $k_1=p, k_2=p-k$. Integrating out the delta functions
\be
\langle J_a(x,u)J_b(y,v)\rangle
 &=&{-}\int \frac {d^dk}{(2\pi )^d}\, e^{ik\cdot(x-y)}\int \frac {d^dp}{(2\pi
)^d}\, {\delta\left(u-\frac {2p-k}2\right)\delta\left(v-\frac
{2p-k}2\right)} \0\\
&&\times \tr \left( \gamma_a\frac {{i}}{{\slashed p}+m}
\gamma_b\frac
{{i}}{{\slashed p-\slashed k}+m}\right)\label{2pt2}
\ee
In components
\be
&&\langle J^{(s)}_{a\mu_1\ldots \mu_{s-1}} (x)J^{(r)}_{b\nu_1\ldots
\nu_{r-1}} (y)\rangle\0\\
&=&{-
{\frac {1}{(s-1)!(r-1)!}}}\int \frac{d^du}{(2\pi)^d}\,
\frac{d^dv}{(2\pi)^d}\, \langle J_a(x,u)J_b(y,v)\rangle \, u_{\mu_1} \ldots
u_{\mu_{s-1}}v_{\nu_1} \ldots {v}_{\nu_{r-1}}\label{2pt3}\\
&=&{-\frac
{1}{2^{s+r-2}(s-1)!(r-1)!}}\int\frac{d^dp}{(2\pi)^d}\int
\frac {d^dk}{(2\pi )^d}\, e^{ik\cdot(x-y)} {(2p-k)_{\mu_1}\ldots
(2p-k)_{\mu_{s-1}}}\0\\
&&\times{(2p-k)_{\nu_1}\ldots (2p-k)_{\nu_{r-1}}}\,\tr \left( \gamma_a\frac
{{i}}{{\slashed p}+m} \gamma_b \frac
{{i}}{{\slashed p-\slashed k}+m}\right)\0
\ee
{Fourier transforming  the above amplitude gives
(\ref{bubble}).}
\subsubsection{3-point function}

The connected 3-pt function for master currents is
\be
&&\langle J_{a_1}(x_1,u_1)  J_{a_2}(x_2,u_2) J_{a_3}(x_3,u_3)\rangle =
\frac {{1}}{3!}\int d^dz_1 \, e^{iz_1\cdot u_1} \frac
{\delta}{\delta
j\left(x_1+\frac {z_1}2\right)}\gamma_{a_1} \frac{\delta}{\delta \overline
j\left(x_1-\frac {z_1}2\right)}\label{3pt0}\\
&&\times \int d^dz_2 \, e^{iz_2\cdot u_2} \frac {\delta}{\delta j\left(x_2+\frac
{z_2}2\right)}\gamma_{a_2} \frac{\delta}{\delta \overline j\left(x_2-\frac
{z_2}2\right)}
\int d^dz_3 \, e^{iz_3\cdot u_3} \frac {\delta}{\delta j\left(x_3+\frac
{z_3}2\right)}\gamma_{a_3} \frac{\delta}{\delta \overline j\left(x_3-\frac
{z_3}2\right)}\0\\
&&\times \int d^dy_1 d^d \bar y_1\int d^dy_2 d^d \bar y_2\int d^dy_3 d^d \bar
y_3\int\frac {d^dv_1}{(2\pi )^d} \frac {d^dv_2}{(2\pi )^d} \frac {d^dv_3}{(2\pi
)^d}  d^dt_1 d^dt_2  d^dt_3   e^{it_1\cdot v_1} e^{it_2\cdot v_2} e^{it_3\cdot
v_3}\0\\
&&\times \overline j\left(y_1-\frac {t_1}2\right)  P_{1,\bar 1}\,j\left(\bar
y_1+\frac {t_1}2\right)
\overline j\left(y_2-\frac {t_2}2\right)  P_{2,\bar 2}\,j\left(\bar y_2+\frac
{t_2}2\right)
\overline j\left(y_3-\frac {t_3}2\right)  P_{3,\bar 3}\,j\left(\bar y_3+\frac
{t_3}2\right)\0
\ee
where
\be
{ P}_{i,\bar i} = {i}\int \frac{d^dk_i}{(2\pi
)^d}\frac{e^{ik_i\cdot(y_i-\bar
y_i-t_i)}}{{\slashed k_i}+m},\quad\quad  i=1,2,3
\label{propxy112233}
\ee
Proceeding as before one finds
\be
&&\langle J_{a_1}(x_1,u_1)  J_{a_2}(x_2,u_2)
J_{a_3}(x_3,u_3)\rangle={i}
\int d^dz_1 \,d^dz_2 \, d^dz_3 \,  e^{i(z_1\cdot u_1+z_2\cdot u_2+z_3\cdot u_3)}
\label{3pt1}\\
&& \times  \int \prod_{i=1}^3\frac{d^dk_i}{(2\pi )^d}e^{ik_1\cdot\left(x_1-x_2 -
\frac {z_1 {+}z_2}2\right)}e^{ik_2\cdot\left(x_2-x_3 - \frac
{z_2 {+}z_3}2\right)}e^{ik_3\cdot\left(x_3-x_1 - \frac
{z_3 {+}z_1}2\right)}\0\\
&& \times \tr \left( \gamma_{a_1}\frac 1{{\slashed k}_1+m} \gamma_{a_2} 
\frac 1{{\slashed k}_2+m} \gamma_{a_3}\frac 1{{\slashed k}_3+m}\right)\0
\ee
Rearranging the terms and defining
\be
k_1=p, \quad\quad k_1-k_2=q_1, \quad\quad k_2-k_3= q_2\0
\ee
$q_1,q_2$ are the momenta of two external outgoing legs. The third has ingoing
momentum $q_1+q_2$. Finally one gets
\be
\langle J_{a_1}(x_1,u_1)  J_{a_2}(x_2,u_2)\!\!\!\!\!&&\!\!\!\!\!
J_{a_3}(x_3,u_3)\rangle = {i} \int \frac{d^dq_1}{(2\pi )^d}
\frac{d^dq_2}{(2\pi )^d}\, e^{i(q_1+q_2)\cdot {x_1}}e^{-i q_1 \cdot {x_2}}
e^{-i q_2 \cdot {x_3}}\label{3pt2}\\
&& \times\,{\delta \left(u_1- \frac {2p-q_1-q_2}2\right)\delta\left(u_2- \frac
{2p-q_1}2\right)\delta\left(u_3- \frac {2p-2q_1-q_2}2\right)} \0\\
&&\times\int \frac{d^dp}{(2\pi )^d} \tr \left({\gamma_{a_1}}\frac 1{{\slashed
p}+m} {\gamma_{a_2} }
\frac 1{{\slashed p -\slashed q}_1+m} {\gamma_{a_3}}\frac 1{{\slashed p
-\slashed q}_1-{\slashed q}_2+m}\right)\0
\ee
To this one must add the cross term. {The Fourier transform of this
amplitude corresponds to (\ref{triangle}).}

\subsubsection{n-point function}

Guess for the n-point function for master currents:
\be
&&\langle J_{a_1}(x_1,u_1) \ldots J_{a_n}(x_n,u_n)\rangle \0\\
&=& -{i^n}\int\frac{d^dp}{(2\pi)^d}\int \prod_{i=1}^{n-1} 
\frac{d^dq_i}{(2\pi )^d} e^{i(q_1+\ldots + q_{n-1})\cdot {x_1}}e^{-i q_1 \cdot
{x_2}}\ldots
e^{-i q_{n-1} \cdot {x_{n}}}\0\\
&&\times \delta \left(u_1- \frac {2p-q_1-\ldots
-q_{n-1}}2\right)\delta\left(u_2- \frac {2p-q_1}2\right)\ldots \delta\left(u_{n}
\frac {2p-2q_1-\ldots -q_{n-1}}2\right)\0\\
&&\times \tr \left( {\gamma_{a_1}}\frac 1{{\slashed p}+m}{ \gamma_{a_2}} 
\frac 1{{\slashed p -\slashed q}_1+m}\ldots {\gamma_{a_{n}}}\frac 1{{\slashed
p -\slashed q}_1-\ldots-{\slashed q}_{n-1}+m}\right)\label{npt}
\ee
$q_1,\ldots, q_{n-1}$ are outgoing external leg momenta, the $n$-th leg has
momentum $q_1+\ldots+ q_{n-1}$. To \eqref{npt} one must add the contributions
with  permutation of $q_1,\ldots, q_{n-1}$.

\vskip 1cm
{Before closing this section let us remark that the integrals
\eqref{2pt3} are of the type already encountered and  explicitly computed in
\cite{BCDGLS,BCDGS}. They can be easily analyzed with the dimensional
regularization. It is likely that also the three-point functions can be
successfully dealt with with analogous techniques, \cite{BoosDavy}.}

\vskip 2cm

{\Large\bf Part II. Relation between the Scalar and Fermion Model}
\vskip 1cm

In the second part of he paper we compare objects and properties of the matter
fermion model with those of the scalar model. The relation between the two
amounts to a quadratic relation between the sources of the former and the
latter.

\section{About HS gauge symmetries}

We compare first the symmetry transformations in the scalar and in the fermion
model.

\subsection{The gauge transformation in the scalar model}
\label{ssec:gtscal}

In the scalar model the action is $S_{(0)}+S_{\rm int}$, where
\be 
S_{0}= \int d^dx \, \partial_\mu \varphi^* \partial^\mu\varphi,\label{S0}
\ee
and 
\be
S_{\rm int}[J, h ]=\sum_{s=0}^\infty \int d^dx \, \frac { 1}{s!} 
J^{(s)}_{\mu_1\ldots\mu_s} (x)  h^{(s)\mu_1\ldots\mu_s} (x) \label{Sintscalar}
\ee
with the currents being chosen in the simple form
\be
J^{(s)}_{\mu_1\ldots\mu_s} (x)  =  {(-)^{s}\frac{i^{s-2}}{2^{s}}}
\sum_{n=0}^s (-1)^n \left( \begin{matrix}s\\n\end{matrix}\right) 
\partial_{(\mu_1}\ldots\partial_{\mu_n} \varphi^*
\partial_{\mu_{n+1}}\ldots\partial_{\mu_s)}\varphi\, , \label{Jsn}
\ee 
The $h$ field is defined by{
\be
&&h(x,u)=\sum_{s=0}^\infty h^{(s)}(x,u)=\sum_{s=0}^\infty \frac 1{s!} 
h_{(s)}^{\mu_1\ldots\mu_s} (x) u_{\mu_1}\ldots u_{\mu_s}\label{hs}
\ee
}
and its transformation is
{
\be
\delta_{\epsilon} h(x,u)= (u\!\cdot\! \partial_x)\epsilon (x,u) -\frac i2
[
h(x,u)
\stackrel{\ast}{,} \epsilon (x,u)]\, , \label{deltah}
\ee
}
{ where the $\ast$-product is defined by
\be
\alpha(x,u)\ast \beta(x,u)=\alpha(x,u)e^{\frac{i}2(\overleftarrow\partial_x
\cdot \overrightarrow\partial_u-\overrightarrow\partial_x \cdot
\overleftarrow\partial_u)} \beta(x,u)\,.
\ee }

More explicitly, representing
\be
h(x,u)=\Phi(x)+ a^\mu(x) u_\mu + \frac 12 h^{\mu\nu}(x)  u_\mu u_\nu+ \frac 16
b^{\mu\nu\lambda}(x) u_\mu u_\nu u_\lambda+\frac 1{4!}
d^{\mu\nu\lambda\rho}(x)u_\mu u_\nu u_\lambda u_\rho + \ldots
\label{hs'}
\ee
and
{
\be
{\eta}(x,u)=
\epsilon(x)+ \xi^\mu(x) u_\mu + \frac 12 \Lambda^{\mu\nu}(x) u_\mu
u_\nu+\frac 1{3!} \Sigma^{\mu\nu\lambda}(x)u_\mu u_\nu 
u_\lambda+\frac 1{4!} P^{\mu\nu\lambda\rho}(x)u_\mu u_\nu 
u_\lambda u_\rho+\ldots
\label{epsilonxn'}
\ee
}
we get the transformations
\be
&&\delta^{(0)} \Phi=0 \label{delta0}\\
&& \delta^{(0)} {a}^\mu= \partial^\mu \epsilon\0\\
&&\delta^{(0)} h^{\mu\nu} = \partial^\mu \xi^\nu +\partial^\nu\xi^\mu
\0\\
&&\delta^{(0)}b^{\mu\nu\lambda} = \partial^\mu\Lambda^{\nu\lambda} +
\partial^\nu\Lambda^{\mu\lambda} + \partial^\lambda
\Lambda^{\mu\nu}\label{delta0b}
\ee
{where $\partial^\mu = \eta^{\mu\nu} \partial_\nu$}, 
which comes from the first term in the RHS of \eqref{deltah}, and
{
\be
\delta^{(1)} \Phi&=&\frac12 \xi\!\cdot\!\partial \Phi
- \frac12 {a}\!\cdot\! \partial\epsilon
+\frac{1}{48}
b^{\nu_{1}\nu_{2}\nu_{3}}\partial_{\nu_{1}}\partial_{\nu_{2}}\partial_{\nu_{3}}
\epsilon
-\frac{1}{48}\Sigma^{\nu_{1}\nu_{2}\nu_{3}}\partial_{\nu_{1}}\partial_{
\nu_{2}}\partial_{\nu_{3}}\Phi
\label{delta1}\\
& &
-\frac{1}{16}\partial_{\nu_{1}}h^{\nu_{2}\nu_{3}}\partial_{\nu_{2}}
\partial_{\nu_{3}}\xi^{\nu_{1}}+\frac{1}{16}\partial_{\nu_{1}}\partial_{\nu_{2}}
 {a}^{\nu_{3}}\partial_{\nu_{3}}\Lambda^{\nu_{1}\nu_{2}} 
\0\\
&  &
-\frac{1}{384}\partial_{\nu_{1}}\partial_{\nu_{2}}b^{\nu_{3}\nu_{4}\nu_{5}} 
\partial_{\nu_{3}}\partial_{\nu_{4}}\partial_{\nu_5}\Lambda^{\nu_{1}\nu_{2}}
+ \frac{1}{384}\partial_{\nu_{1}}\partial_{\nu_{2}}\Sigma^{\nu_{3}\nu_{4}
\nu_{5}}\partial_{\nu_{3}}\partial_{\nu_{4}}\partial_{\nu_5}h^{\nu_{1}\nu_{2}}
\0\\
\delta^{(1)}  {a}^\mu&=&{\frac12} \xi\!\cdot\!\partial  {a}^\mu
+{\frac12} \partial_\rho
\Phi\Lambda^{\mu\rho} {-\frac12} \partial_\rho \epsilon
\,h^{\mu\rho} 
- {\frac12} {a} \!\cdot\!\partial \xi^\mu\0\\
 & 
&+\frac{1}{48}b^{\nu_{1}\nu_{2}\nu_{3}}\partial_{\nu_{1}}\partial_{
\nu_{2}}\partial_{\nu_{3}}\xi^{\mu}
-\frac{1}{48}\Sigma^{\nu_{1}\nu_{2}\nu_{3}}\partial_{\nu_{1}}\partial_{
\nu_{2}}\partial_{\nu_{3}} {a}^{\mu}
\0\\&&
-\frac{1}{16}\partial_{\nu_{1}}
b^{\nu_{2}\nu_{3}\mu}\partial_{\nu_{2}}\partial_{\nu_{3}}\xi^{\nu_{1}}
+\frac{1}{16}\partial_{\nu_{1}}
\Sigma^{\nu_{2}\nu_{3}\mu}\partial_{\nu_{2}}\partial_{\nu_{3}} {a}^{\nu_{1}}
\0\\
  &  &{
-\frac{1}{16}\partial_{\nu_{1}}{h}^{\nu_{2}\nu_{3}}\partial_{\nu_{2}}\partial_{
\nu_{3}}\Lambda^{\nu_{1}\mu}+\frac{1}{16}\partial_{\nu_{1}}\partial_{\nu_{2}
}{h}^{\nu_{3}\mu}\partial_{\nu_{3}}\Lambda^{\nu_{1}\nu_{2}}} \0\\
&&
-\frac{1}{384}
\partial_{\nu_{1}}\partial_{\nu_{2}}\partial_{\nu_{3}}\Sigma^{\nu_{4}\nu_{5}\mu}
\partial_{\nu_{4}}\partial_{\nu_5}b^{\nu_{1}\nu_{2}\nu_{3}}
+\frac{1}{384}
\partial_{\nu_{1}}\partial_{\nu_{2}}\partial_{\nu_{3}}b^{\nu_{4}\nu_{5}\mu}
\partial_{\nu_{4}}\partial_{\nu_5}\Sigma^{\nu_{1}\nu_{2}\nu_{3}}
\0\\
\delta^{(1)} h^{\mu\nu} &=&{\frac12}\xi\!\cdot\!\partial
h^{\mu\nu}
-\partial_\rho\xi^{\left(\mu\right.} h^{\nu)\rho} 
+\partial_\rho  {a}^{(\mu} \Lambda^{\nu)\rho}   
-{\frac12} {a}\!\cdot\! \partial \Lambda^{\mu\nu} 
{-\frac12}\partial_\lambda \epsilon \,b^{\lambda\mu\nu}
+ \frac12 \Sigma^{\mu\nu\rho}\partial_\rho \Phi
\0\\
 &  &
+\frac{1}{48}b^{\nu_{1}\nu_{2}\nu_{3}}\partial_{\nu_{1}}\partial_{
\nu_{2}}\partial_{\nu_{3}}\Lambda^{\mu\nu}
-\frac{1}{96}\Sigma^{\nu_{1}\nu_{2}\nu_{3}}\partial_{\nu_{1}}\partial_{
\nu_{2}}\partial_{\nu_{3}}h^{\mu\nu}
\0\\&&
-\frac18\partial_{\nu_{1}}b^{
\nu_{2}\nu_{3}\left(\mu\right.}\partial_{\nu_{2}}\partial_{\nu_{3}}\Lambda^{
\left.\nu\right)\nu_{1}}
+\frac18\partial_{\nu_{1}}\Sigma^{
\nu_{2}\nu_{3}\left(\mu\right.}\partial_{\nu_{2}}\partial_{\nu_{3}}h^{
\left.\nu\right)\nu_{1}}
\0\\&&
+\frac{1}{16}\partial_{\nu_{1}}\partial_{\nu_{2}}b^{\nu_{3}\mu\nu}\partial_{\nu_
{3}}\Lambda^{\nu_{1}\nu_{2}}
-\frac{1}{16}\partial_{\nu_{1}}\partial_{\nu_{2}}\Sigma^{\nu_{3}\mu\nu}\partial_
{\nu_{3}}h^{\nu_{1}\nu_{2}}
\0\\
\delta^{(1)} b^{\mu\nu\lambda} &=& {\frac12} \xi\!\cdot\!\partial
b^{\mu\nu\lambda}
-{\frac32} \partial_\rho \xi^{(\mu} b^{\nu\lambda)\rho}
+{\frac32} \partial_\rho h^{(\mu\nu} \Lambda^{\lambda)\rho}
-{\frac32}  h^{\rho(\mu} \partial_\rho \Lambda^{\nu\lambda)} 
\0\\&&
- \frac12 {a}^\rho \partial_\rho \Sigma^{\mu\nu\lambda}
+\frac32 \Sigma^{\rho(\mu\nu}\partial_\rho  {a}^{\lambda)}
\0\\&&
-\frac1{96} \Sigma^{\nu_{1}\nu_{2}\nu_{3}} 
 \partial_{\nu_{1}}\partial_{\nu_{2}}\partial_{\nu_{3}} b^{\mu\nu\lambda}
+\frac1{96} b^{\nu_{1}\nu_{2}\nu_{3}} 
 \partial_{\nu_{1}}\partial_{\nu_{2}}\partial_{\nu_{3}} \Sigma^{\mu\nu\lambda}
\0\\&&
-\frac3{16} \partial_{\nu_{1}} b^{\nu_{2}\nu_{3}(\mu} 
 \partial_{\nu_{2}}\partial_{\nu_{3}} \Sigma^{\nu\lambda)\nu_{1}}
+\frac3{16} \partial_{\nu_{1}} \Sigma^{\nu_{2}\nu_{3}(\mu} 
 \partial_{\nu_{2}}\partial_{\nu_{3}} b^{\nu\lambda)\nu_{1}}
\0
\ee
}
which is the full contribution for the gauge transformations of spin 1, 2 and 3,
under the assumption that all fields of spin higher than 3 are disregarded.

These transformations suggest that the fields $\Phi$, {$a^\mu$},
$h^{\mu\nu}$ do not
coincide with the standard scalar vector and metric fields $\tilde\Phi$, $\tilde
A^\mu$, $\tilde g^{\mu\nu}$. In fact restricting to  U(1) gauge and diff
transformations, they transform as \eqref{delta0} and
\be
\delta^{(1)} \Phi&=&{\frac12} \xi\!\cdot\!\partial \Phi -{\frac12} {a}\!\cdot\!
\partial\epsilon
 - {\frac{1}{16}\partial_{\mu}h^{\nu\rho}\, \partial_{\nu}
\partial_{\rho}\xi^{\mu}}
\label{restricted0} \\
\delta^{(1)}  {a}^\mu&=&{\frac12} \xi\!\cdot\!\partial {a^\mu}
 {-\frac12} \partial_\rho \epsilon \,h^{\mu\rho} - {\frac12} {a}
\!\cdot\!{\partial\xi^\mu}
\label{restricted1} \\
\delta^{(1)} h^{\mu\nu} &=&{\frac12}\xi\!\cdot\!\partial
h^{\mu\nu}-{\frac12}\partial_\rho
\xi^\mu h^{\rho\nu} -{\frac12}\partial_\rho \xi^\nu h^{\rho\mu} 
\label{restricted2}
\ee
The standard fields instead transform as
\be
{\delta^{(1)}} \tilde\Phi  &=& \zeta\!\cdot\! \partial
\tilde{\Phi} 
\label{standard0} \\
{\delta^{(1)}} \tilde  {{a}}^\mu &=&  \zeta\!\cdot\! \partial
\tilde  {{a}}^\mu 
- \partial_\lambda \zeta^\mu \tilde {{a}}^\lambda
\label{standard1} \\ 
{\delta^{(1)}} \tilde g^{\mu\nu} &=& \zeta\!\cdot \!\partial
\tilde g^{\mu\nu}-
\partial_\lambda \zeta^\mu \tilde g^{\lambda \nu}- \partial_\lambda \zeta^\nu
\tilde g^{\mu\lambda}
\label{standard2}
\ee
We can reproduce the above transformations if we identify the fields
as follows
\be
\Phi&=& -\frac 14 \tilde  {{a}}_\mu \tilde  {{a}}^\mu -{\frac 1
{32} \left( \partial_\mu \tilde h_{\nu\rho} \partial^\nu \tilde h^{\mu\rho}
-\frac 12 \partial_\mu \tilde h_{\nu\rho}  \partial^\mu \tilde
h^{\nu\rho}\right)}+ \ldots 
\label{phiphitilde}\\
{ {a}}^\mu&=& \tilde g^{\mu\nu} \tilde {{a}} _\nu = \tilde  {{a}}^\mu - \frac 12
\tilde
h^{\mu\nu} \tilde  {{a}}_\nu
\label{AAtilde} \\
g^{\mu\nu} &\equiv& \eta^{\mu\nu}-\frac 12   h^{\mu\nu}=\eta^{\mu\nu}-\frac 12 
\tilde h^{\mu\nu}
\label{ggtilde}
\ee
together with $\xi^\mu= 2\, \zeta^\mu$. { In \eqref{phiphitilde}
ellipses stand for terms of order 3 and higher in $\tilde h$. For the second
term in the RHS reproduces the third term in the RHS of \eqref{restricted0} up
to terms quadratic in the field $\tilde h$. In fact a series in $\tilde h$ is
required to reproduce \eqref{restricted0} exactly. In Sec.\ref{ssec:connge} we
shall show that a unique term can replace this series, but for this it will be
necessary to introduce the vielbein.}

Due to the above transformation properties it is natural to refer to the master
field $h$ as a {\it metric-like field} and to the HS geometry resulting from
integrating out scalar matter, as metric-like geometry.

\subsection{The HS gauge transformation in the fermion model}
\label{s:gaugetransf}

Let us come now to the HS gauge transformation from the fermion model. It
comes from the
invariance of the fermion action written in the operator form
\be
S= \langle \overline \psi | - \gamma \!\cdot\! (\widehat P- \widehat H) -m
|\psi\rangle \label{Sop}
\ee   
where $\widehat P_\mu$ is the momentum operator whose symbol is the classical
momentum $u_\mu$ and $\widehat H$ is an operator whose symbol is $h(x,u)$. 
For \eqref{Sop} is trivially invariant under the operation
\be
S= \langle \overline \psi | {\widehat O}  {\widehat O}^{-1} \widehat G{\widehat
O}{\widehat O}^{-1}|\psi\rangle 
\ee
where $\widehat G=  - \gamma \!\cdot\! (\widehat P- \widehat H) -m$. So it is
invariant under
\be
\widehat G \longrightarrow {\widehat O}^{-1} \widehat G{\widehat O},\quad\quad
|\psi\rangle \longrightarrow {\widehat O}^{-1}|\psi \rangle\label{G'}
\ee
Writing $\widehat O = e^{-i \widehat E}$ we find the infinitesimal
version.
\be
\delta |\psi \rangle = i \widehat E |\psi\rangle,\quad\quad \delta \langle
\overline \psi| =- i  \langle \overline \psi| \widehat E, \label{deltapsi}
\ee
Passing from operators to symbols, the transformation of $h^\mu(x,u)$ is made of
two pieces,  
\be
\mathrm{Symb}\big( [\gamma\!\cdot \!\widehat P, \widehat E]\big)=
 [\gamma\!\cdot\! u \stackrel{\ast}{,} \varepsilon(x,u)]
= -i \gamma\!\cdot\!\partial_x  \varepsilon(x,u)
\label{PE'}
\ee
and
 \be
\mathrm{Symb}\big([\gamma \!\cdot \!\widehat H,\widehat E]\big) =
 [\gamma\!\cdot \! h (x,u) \stackrel{\ast}{,} \varepsilon(x,u)]   \label{HE}
\ee
In these equations $\gamma \!\cdot \!\widehat H= \gamma^a\widehat H_a$, where
$a$ will be understood as a
flat index. Of course, since the background is flat, writing $\gamma^a H_a$ or
$\gamma^\mu H_\mu$ is the same, but the first writing leads to the correct
interpretation. So, from now on, we will use $h_a(x,u)$.  In the sequel the
index $a$ will play a special role. In terms of symbols, we thus rewrite
\eqref{deltahxp} as 
\be
\delta_\varepsilon h_a(x,u) = \partial_a^x 
\varepsilon(x,u)-i [h_a(x,u) \stackrel{\ast}{,} \varepsilon(x,u)] 
\equiv {\cal D}^{x\ast}_a \varepsilon(x,u)\label{deltahxpcov}
\ee

Next, it is helpful to see all the above formulas in components.  To avoid a
proliferation of indices, let us write the expansion of $h_a(x,u)$ as 
\be
h_a(x,u) &=& A_a(x) +\chi_a^\mu(x) u_\mu 
+ \frac 12 b_a^{\mu\nu} u_\mu u_\nu+\frac 16 c_a^{\mu\nu\lambda} u_\mu u_\nu
u_\lambda
+ \frac 1{4!}d_a^{\mu\nu\lambda\rho}u_\mu u_\nu u_\lambda
u_\rho\0\\
&&+\frac 1{5!} f_a^{\mu\nu\lambda\rho\sigma}u_\mu u_\nu u_\lambda
u_\rho u_\sigma+\ldots\label{haunn}
\ee
Notice that in the expansion \eqref{hmmm} the indices $\mu_1,\ldots,\mu_n$ are
upper (contravariant), as it should be, because in the Weyl quantization
procedure the momentum has lower index, since it must satisfy $[x^\mu, p_\nu] =i
\,\delta_\nu^\mu$. Of course when the background metric is flat  the indices $a$
and $\mu_i$ are on the same footing, but it is useful to keep them distinct.

Similarly we write
\be
\varepsilon(x,u)= \epsilon(x) +\xi^\mu u_\mu+\frac 12 \Lambda^{\mu\nu}u_\mu
u_\nu+\frac 1{3!} \Sigma^{\mu\nu\lambda}u_\mu
u_\nu u_\lambda+\frac 1{4!} P^{\mu\nu\lambda\rho}u_\mu
u_\nu u_\lambda u_\rho+\frac 1{5!}\Omega\!\cdot\! u^5+\ldots\label{epsxu}
\ee
To avoid a proliferation of symbols we use for the component of
$\varepsilon(x,u)$ the same symbols as 
for the expansion of $\eta(x,u)$ in the scalar model. As we shall see they are
not the same (in fact $\eta=2\varepsilon$). We invite the reader to remember the
distinction.

The transformation \eqref{deltahxpcov} reads. to lowest order,
\be
&& \delta^{(0)} A_a= \partial_a \epsilon\0\\
&&\delta^{(0)} \chi_a^{\nu} = \partial_a \xi^\nu  \0\\
&&\delta^{(0)}b_a{}^{\nu\lambda} = \partial_a\Lambda^{\nu\lambda}
\label{deltaAhb}
\ee

To first order we have
\be
\delta^{(1)} A_a &=& \xi\!\cdot\!\partial A_a - \partial_\rho \epsilon
\,\chi_a^{\rho} \label{delta1Ahb}\\
\delta^{(1)} \chi_a^{\nu} &=& \xi\!\cdot\!\partial \chi_a^\nu-\partial_\rho
\xi_\nu \chi_a^\rho 
+ \partial^\rho A_a \Lambda_{\rho}{}^\nu   - \partial_\lambda \epsilon
\,b_a{}^{\lambda\nu}\0\\
\delta^{(1)} b_a^{\nu\lambda} &=& \xi\!\cdot\!\partial b_a{}^{\nu\lambda}
-\partial_\rho \xi^\nu b_a{}^{\rho\lambda}- \partial_\rho \xi^\lambda 
b_a{}^{\rho\nu }+\partial_\rho \chi_a^{\nu} \Lambda^{{\rho\lambda}}
+\partial_\rho
\chi_a^{\lambda} \Lambda^{{\rho\nu}}
- \chi_a^{\rho} \partial_\rho \Lambda_{\nu\lambda} \0
\ee
The next nontrivial order contains terms with three derivatives, and so on.

Let us denote now by $\tilde A_a$ the standard U(1) gauge field and by 
$\tilde e_a^\mu=\delta_a^\mu -\tilde \chi_a^\mu$ the standard inverse vielbein,
and let us restrict to gauge and
diff transformations alone, we have
\be
\delta \tilde A_a&\equiv& \delta \left(\tilde e_a^\mu \tilde A_\mu\right)\equiv
\delta  \left((\delta_a^\mu -\tilde \chi_a^\mu) \tilde
A_\mu\right)\label{standardtransf}\\
&=&\left(-\xi\!\cdot\!\partial \tilde \chi_a^\mu +\partial_\lambda \xi^\mu
\tilde \chi_a^\lambda\right) \tilde A_\mu+(\delta_a^\mu -\tilde
\chi_a^\mu)\left(\partial_\mu\epsilon + \xi\!\cdot\! \tilde A_\mu\right)\approx
\partial_a\epsilon + \xi\!\cdot\! \tilde A_a- \tilde \chi_a^\mu
\partial_\mu\epsilon \0
\ee
and
\be
\delta \tilde e_a^\mu \equiv  \delta  (\delta_a^\mu -\tilde \chi_a^\mu) =
\xi\!\cdot\! \partial\tilde e_a^\mu -\partial_\lambda \xi^\mu \tilde e_a^\lambda
= -
\xi\!\cdot\! \tilde \chi_a^\mu -\partial_a \xi^\mu +\partial_\lambda \xi^\mu
\tilde \chi_a^\lambda\label{deltaeamu}
\ee
so that
\be
\delta \tilde \chi_a^\mu= \xi\!\cdot\! \partial\tilde \chi_a^\mu +\partial_a
\xi^\mu
-\partial_\lambda \xi^\mu \tilde \chi_a^\lambda\label{deltaeamu1}
\ee
where we have retained only the terms at most linear in the fields.
From the above we see that we can make the identifications
\be
A_a= \tilde A_a, \quad\quad \chi_a^\mu = \tilde \chi_a^\mu\label{identAAee}
\ee

The transformations \eqref{deltaAhb}, \eqref{delta1Ahb} are consistent with
Riemannian geometry. Concerning the HS theory, it contains more
than symmetric tensors: beside the completely symmetric   $h^{a\mu_1\ldots
\mu_n}$ it contains also a Lorentz representation in 
which the index $a$ and one of the other indices are antisymmetric.

{This point deserves a further comment. The interpretation of $_a$ in 
$h_a{}^{\mu_1\ldots \mu_n}$ as a flat index almost calls for local Lorentz 
invariance in the effective action (in this regard see for instance 
\cite{Lopatin}). This question is discussed in detail in II, \cite{BCDGSII}. It 
is shown there that the action \eqref{S} does admit a local Lorentz invariant 
extension. As a consequence it is expected that the relevant Ward identities, 
barring anomalies, are obeyed, leading to Lorentz invariant effective actions. 
In II explicit examples are constructed. One of the consequences of this 
symmetry is that the antisymmetric part of $\chi_a^\mu$ can be identified with 
local Lorentz gauge parameters and thus eliminated by gauge fixing. As for the 
HS components of  $h_a{}^{\mu_1\ldots \mu_n}$ the problem is open. For instance 
in II it is shown that, in the YM-like models, in order to guarantee the 
existence of the relevant propagators,  $h^{a\mu_1\ldots \mu_n}$ must be 
traceless in the $\mu_i$ indices. In general, however, this problem needs to be 
further investigated.
}

\subsection{Analogy with gauge transformations in gauge theories}

Notice first that, in eq.\eqref{deltahxpcov} and \eqref{deltaAhb}, the
derivative $\partial_a$
means 
\be
\partial_a = \delta_a^\mu \partial_\mu,\label{partialsuba}
\ee
not
\be
\partial_a = e_a^\mu \partial_\mu= \left(e_a^\mu
-\chi_a^\mu+\ldots\right)\partial_\mu,\label{partialsuba'}
\ee
for the linear correction $ -\chi_a^\mu\partial_\mu$ is
contained in the term $ -i [h_a(x,u) \stackrel{\ast}{,} \varepsilon(x,u)]$, see
for instance the second term in the RHS of the first equation
\eqref{delta1Ahb}. 
Therefore the transformation \eqref{deltahxpcov} looks similar to
an ordinary gauge transformation of a non-Abelian gauge field
\be
\delta_\lambda A_a = \partial_a \lambda + [A_a, \lambda]\label{gaugetransf}
\ee
where $A_a= A_a^\alpha T^\alpha, \lambda = \lambda^\alpha T^\alpha$, $T^\alpha$
being the Lie algebra generators.

In gauge theories it is useful to represent the gauge potential as a connection
one form ${\bf A}= A_a dx^a$, so that \eqref{gaugetransf} becomes
\be
\delta_\lambda {\bf A} = {\bf d} \lambda + [{\bf A},
\lambda]\label{gaugetransform}
\ee
We can do the same for \eqref{deltahxpcov}
\be
\delta_\varepsilon {\bf h}(x,u) ={\bf d} 
\varepsilon(x,u)-i [{\bf h} (x,u) \stackrel{\ast}{,} \varepsilon(x,u)] \equiv 
\bfD^\ast \varepsilon (x,u) \label{deltahxpbf}
\ee
where ${\bf d}= \partial_a\, dx^a, {\bf h}= h_a dx^a$ and $x^a$ are coordinates
in the tangent spacetime, and it is understood that 
\be
  [{\bf h} (x,u) \stackrel{\ast}{,} \varepsilon(x,u)]=  [h_a (x,u)
\stackrel{\ast}{,} \varepsilon(x,u)]dx^a\0
\ee

Like in gauge theories it is straightforward to introduce the curvature 
\be
G_{ab}= \partial_a h_b - \partial_b h_a -i [h_a \stackrel{\ast}{,} h_b ]
\label{Gab}
\ee
whose transformation rule is
\be
\delta_\varepsilon G_{ab}=-i [G_{ab}\stackrel{\ast}{,}
\varepsilon],\label{deltaGab}
\ee
as well as the curvature two-form
\be
{\bf G} = {\bf d} {\bf h}
 -\frac i 2 [ {\bf h}\stackrel {\ast}{,}{\bf h}],\label{curv1} 
\ee
with the transformation property
\be
\delta_\varepsilon {\bf G} = -i [{\bf G}  \stackrel {\ast}{,}\varepsilon]
\label{deltaG}
\ee
In part III of this paper this formalism will be applied to the construction of
anomalies  and in paper II  to the construction of Yang-Mills and
Chern-Simons-like theories.

\smallskip

The YM-like interpretation presented so far for $h_a$ and relevant
constructs, using the analogy with gauge theories, is not the only possible one.
There is another, which we have already dubbed {\it geometry-like}, which will
be developed in the paper III.

\subsection{Connection between metric-like and frame-like master fields}
\label{ssec:connge}

It is expected that there is a common set of higher spin fields, which couples
both to the fermionic and bosonic matter, so it is important to find the
connection
between frame-like and metric like fields. One can show that if one defines the
composite master field\footnote{In the forthcoming paper III we shall show that
expression (\ref{hmhfc}) naturally follows from the geometry-like formalism we
develop there.}
\be \label{hmhfc}
h(x,u) = 2\, u_a h^a(x,u) - h_a(x,u) \ast h^a(x,u) 
\ee
and calculates its HS transformation by using (\ref{deltahxpcov}), one gets
\be
\delta_\varepsilon h(x,u) = 2\, (u \cdot \partial^x)\, \varepsilon(x,u)
 - i\, [\,h(x,u) \stackrel{\ast}{,} \varepsilon(x,u)]
\ee
We see that it becomes identical to (\ref{deltah}) if we make the
identifications
\be
\eta(x,u) = 2\, \varepsilon(x,u) \label{epsilo2varepsilon}
\ee
The expression (\ref{hmhfc}) defines the simplest metric-like master field
constructed from the frame-like master field which satisfies the transformation
law
(\ref{deltah}), and so can be used to ``linearly" couple HS fields to bosonic
matter. It is now obvious that coupling to bosonic matter is at least quadratic
in powers of (fundamental) HS fields.

By using (\ref{hmhfc}) we can now express metric-like spacetime fields in terms
of frame-like spacetime fields by Taylor expanding both sides of the equation
around $u=0$. For
example, the scalar component is given by
\be
\Phi(x) \!\!\!&=&\!\!\! - h_a(x,u) \star h^a(x,u) \Big|_{u=0} 
\0 \\
\!\!\!&=&\!\!\! - A_a(x)A^a(x){ - }{\frac{1}{2} \partial_\mu
\chi_a{}^\nu \partial_\nu \chi^{a\mu}} + \ldots
\label{Phiae}
\ee
{where ellipses denote terms containing fields of spin $s\ge3$.}
For spin 1 and 2 components one gets
{
\be
a^\mu(x) \!\!\!&=&\!\!\! 2\, e_a{}^\mu(x) A^a(x) + \ldots 
\label{AmuAa}\\
h^{\mu\nu}(x) \!\!\!&=&\!\!\! 2 \big(\delta^\mu_a \chi^{a\nu}(x) + \delta^\nu_a
\chi^{a\mu}(x)
 - \chi^{a\mu}(x) \chi_a{}^\nu(x) \big) + \ldots
\label{hchi}
\ee
where 
\be
e_a{}^\mu(x) \equiv \delta_a^\mu - \chi_a{}^\mu(x)
\ee
The relations \eqref{AmuAa} and \eqref{hchi} are natural once we recognize
$e_a{}^\mu(x)$ as vielbein and
\be
g^{\mu\nu}(x) = \eta^{\mu\nu} - \frac{1}{2} h^{\mu\nu}(x)
\ee
as metric. The relation \eqref{hchi} then becomes
\be
g^{\mu\nu}(x) = \eta^{ab} e_a{}^\mu(x)\, e_b{}^\nu(x)
\ee
which is the standard relation between (inverse) metric and vielbein.}

As for \eqref{Phiae}, while the first term produces the standard ``seagull" term
in the Klein-Gordon coupling to the $U(1)$ gauge field, the second term is of
more
mysterious nature.  Superficially it does not look locally
Lorentz invariant, however,
as we noticed in our comment after eq.\eqref{ggtilde} it is likely to reproduce
an infinite series 
in $h_{\mu\nu}$, and so to be local Lorentz invariant. Incidentally this tells
us that the question of local Lorentz invariance in this formalism is rather
subtle, see II.
It also suggests that the coupling of Klein-Gordon field to HS fields apparently
cannot be described in terms of Riemannian geometry. We shall say more on this
in the paper III.

\vskip 2cm

\vskip 2cm

{\Large \bf Part III. Chern-Simons terms and Anomalies}
\vskip 1cm

The HS gauge transformation \eqref{deltahxp} suggests  in an obvious way an
analogy with gauge transformations in ordinary (non-Abelian) gauge theories. In
both cases they are realized via Ward identities  on the physical amplitudes.
But as an effect of quantization the Ward identities in a second quantized
theory may be violated, or anomalous. It is to be expected that a similar
possibility exists also in the matter theories coupled to external HS sources
introduced in this paper. In the third part we analyse the form of the possible
obstructions that may appear in the Ward identities in such context.

\section{Obstructions}

In theory we can compute  $\EW^{(n)}$ by means of formula (2.52) of
\cite{BCDGPS} 
or by means of Feynman diagrams. In practice we may find obstacles. 

The first is the possibility that the one-point function does not vanish:
$\EW^{(1)}\neq 0$.
In this case \eqref{WI1} is modified and the natural setting is a {\it curved}
$L_\infty$.

The second is that the results one obtains may
not satisfy the WI's. If it is so, however, we are helped by the consistency
conditions. For, if
\be
\delta_\varepsilon \EW[h] = \EA[\varepsilon, h]\neq 0, \label{A1}
\ee
as a consequence of 
\be 
\left(\delta_{\varepsilon_2 }\delta_{\varepsilon_1 }- \delta_{\varepsilon_1
}\delta_{\varepsilon_2}\right) h^\mu(x,u) &=&  i\left( \partial_x
[{\varepsilon_1
}\stackrel{\ast}{,}{\varepsilon_2 }](x,u) -i [h^\mu(x,u)\stackrel{\ast}{,} 
[{\varepsilon_1 }\stackrel{\ast}{,}{\varepsilon_2 }](x,u) ]] \right)
\0 \\
&=&   i\, {\cal D}^{\ast\mu}_x [{\varepsilon_1}\stackrel{\ast}{,}{\varepsilon_2
}](x,u),\label{e1e2}
\ee
we must have
\be
\delta_{\varepsilon_2}\EA[\varepsilon_1, h]-\delta_{\varepsilon_1}
\EA[\varepsilon_2, h]=
\EA[[\varepsilon_1\stackrel{\ast}{,} \varepsilon_2],h]\label{A2}
\ee
The simplest possibility we may encounter is that $\EA[\varepsilon, h]$ is
trivial, i.e.
\be
\EA[\varepsilon, h]= \delta_\varepsilon \EC[h],\label{C1}
\ee
{where $\EC[h]$ is an integrated local counterterm. 
In such a case} we can recover invariance in the form
\be
\delta_\varepsilon \left(  \EW[h]- \EC[h]\right)=0\label{A3}
\ee
If however \eqref{C1} is not true for any local choice of $\EC[h]$, then we are
faced
with 
a true anomaly, which breaks the covariance of the effective action.

The case of trivial anomalies is what often occurs in Feynman diagram
calculations
when tadpoles or seagull diagrams are disregarded: the WI is violated, but
invariance can be restored by suitable subtractions of local terms.

When, instead, \eqref{C1} is not true for any local choice of $\EC[h]$, it means
that
there is a 
true obstruction to building a gauge covariant theory. The true anomalies, i.e.
the non-trivial cocycles \eqref{A1}, are the mathematical objects that classify
such obstructions.

We can apply the frame-like formalism to the construction of CS actions and
anomalies.

\subsection{HS CS terms}

The idea in this subsection is to mimic standard constructions of ordinary field
theories in
the framework of HS. For instance, beside \eqref{curv1} we can introduce the
standard 
(ordinary gauge theories) definitions
\be
 {\bf G}_t = {\bf d}
{\bf h}_t
 -\frac i 2 [ {\bf h}_t \stackrel {\ast}{,}{\bf h}_t],\quad\quad {\bf h}_t= t
{\bf h} \label{Gt}
\ee
Let us quote some formulas that will be useful later
\be
\frac {d}{dt} \bfG_t&=& \bfd \bfh -i t[\bfh \stackrel {\ast}{,} \bfh]= \bfd_t
\bfh, \quad\quad
\bfd_t = \bfd -i [\bfh_t\stackrel {\ast}{,}\quad]\label{dGt}\\
\bfd \bfG_t& =& i [\bfh_t\stackrel {\ast}{,}\bfG_t],\quad\quad \delta \bfG_t =
\bfd \delta \bfh_t -i [ \bfh_t \stackrel {\ast}{,} \delta\bfh_t] = \bfd_t \delta
\bfh_t\label{deltaGt}
\ee

The difference in the HS case is that, unlike in ordinary gauge theories, we
cannot 
use graded commutativity.
There is also another difficulty: in the HS case we don't have a trace at our
disposal.
The only object with trace properties we can define is 
\be
\langle\!\langle f\ast g\rangle\!\rangle = \int d^dx \int \frac {d^du}{(2\pi)^d}
f(x,u)\ast g(x,u) = \int d^dx \int \frac {d^du}{(2\pi)^d} f(x,u) g(x,u)=
\langle\!\langle g\ast f\rangle\!\rangle  \label{trace}
\ee
From this, plus associativity, it follows that
\be
&&\langle\!\langle f_1 \ast f_2\ast \ldots \ast f_n\rangle\!\rangle=
\langle\!\langle f_1 \ast (f_2\ast \ldots \ast f_n)\rangle\!\rangle\0\\
&&=(-1)^{\epsilon_1(\epsilon_2+\ldots+\epsilon_n)} \langle\!\langle  (f_2\ast
\ldots \ast f_n)\ast
f_1\rangle\!\rangle=(-1)^{\epsilon_1(\epsilon_2+\ldots+\epsilon_n)} 
\langle\!\langle  f_2\ast \ldots \ast f_n\ast f_1\rangle\!\rangle\label{cycl}
\ee
where $\epsilon_i$ is the Grassmann degree of $f_i$. In particular
\be
\langle\!\langle [f_1 \stackrel{\ast}{,} f_2\ast \ldots \ast
f_n\}\rangle\!\rangle=0\label{comm0}
\ee
where $[\quad  \stackrel{\ast}{,}\quad\}$ is the $\ast$-commutator or
anti-commutator, as appropriate.

But in order to exploit this property we have to integrate over the full phase
space. Therefore it is impossible to reproduce the unintegrated descent
equations like in the ordinary gauge theories. The best we can do is to try to
reproduce each
equation separately in integrated form. So let us start from the phase space
integral with $n$ $\bfG$ entries  
\be
\langle\!\langle \bfG \ast \bfG \ast \ldots\ast \bfG \rangle\!\rangle
\ee
Here $\langle\!\langle\quad \rangle\!\rangle$ means integration over a phase
space of dimension $4n$.
Then consider the expression with $n-1$ $\bfG_t$ entries
\be
&&\int_0^1 dt \langle\!\langle \bfd \left( \bfh \ast \bfG_t \ast \ldots
\ast\bfG_t \right) \rangle\!\rangle= \int_0^1 dt \langle\!\langle \bfd \bfh \ast
\bfG_t \ast \ldots\ast \bfG_t  \rangle\!\rangle\label{cs1}\\
&&\quad\quad-  \int_0^1 dt \langle\!\langle  \bfh \ast \bfd\bfG_t \ast
\ldots\ast \bfG_t  \rangle\!\rangle-\ldots- \int_0^1 dt \langle\!\langle  \bfh
\ast \bfG_t \ast \ldots\ast  \bfd \bfG_t  \rangle\!\rangle\0\\
&=& \int_0^1 dt \Big(\langle\!\langle \bfd \bfh \ast \bfG_t \ast \ldots\ast
\bfG_t  \rangle\!\rangle-   i \langle\!\langle  \bfh \ast
[\bfh_t\stackrel{\ast}{,}\bfG_t ]\ast \ldots\ast \bfG_t 
\rangle\!\rangle-\ldots\0\\
&&\ldots-  i \langle\!\langle  \bfh \ast \bfG_t \ast \ldots\ast 
[\bfh_t\stackrel{\ast}{,}\bfG_t ]\Big)  \rangle\!\rangle
\ee
using the first of \eqref{deltaGt}. Then using the first of \eqref{dGt} together
with \eqref{cycl} (or \eqref{comm0}) one gets
\be
&&\int_0^1 dt \langle\!\langle \bfd \left( \bfh \ast \bfG_t \ast \ldots
\ast\bfG_t \right) \rangle\!\rangle=
\int_0^1 dt \langle\!\langle( \bfd\bfh-i [\bfh_t\stackrel{\ast}{,}\bfh]) \ast
\bfG_t \ast \ldots\ast \bfG_t  \rangle\!\rangle\label{cs2}\\
&=& \int_0^1 dt \langle\!\langle(\frac {d}{dt} \bfG_t   \ast \bfG_t \ast
\ldots\ast \bfG_t  \rangle\!\rangle=
\frac 1n \int_0^1dt \, \frac d{dt} \langle\!\langle \bfG_t   \ast \bfG_t \ast
\ldots\ast \bfG_t  \rangle\!\rangle\0\\
&=& \frac 1n \langle\!\langle \bfG \ast \bfG \ast \ldots\ast \bfG
\rangle\!\rangle\0
\ee
Since they are integrated over a spacetime of dimension $d=2n$ these expressions
vanish (unless the spacetime is topologically nontrivial), but this is the way
we identify the primitive functional action for HS CS terms in dimension
$d=2n-1$:
\be
{\cal CS}(\bfh)=n \int_0^1 dt \langle\!\langle \bfh  \ast \bfG_t \ast \ldots
\ast\bfG_t
 \rangle\!\rangle\label{CSh}
\ee
where $\langle\!\langle\quad \rangle\!\rangle$ means now integration over a
phase space of dimension $4n-2$.
The important thing is to prove that $CS(\bfh)$ is invariant under the HS gauge
transformation \eqref{deltahxpbf}:
\be
\delta_\varepsilon {\cal CS}(\bfh)&=&n \int_0^1 dt\Big( \langle\!\langle
\bfD\varepsilon\ast \bfG_t \ast \ldots \ast\bfG_t  \rangle\!\rangle+t
\langle\!\langle \bfh  \ast  \bfd_t \bfD\varepsilon \ast \ldots \ast\bfG_t 
\rangle\!\rangle+ \ldots\label{deltaCS}\\
&&\dots + t \langle\!\langle \bfh  \ast \bfG_t  \ast \ldots \ast \bfd_t
\bfD\varepsilon \rangle\!\rangle\Big)\0
\ee
Since $\bfd_t \bfG_t=0$ we can collect the symbol $\bfd_t$ and integrate by
parts in space to obtain
\be
\delta_\varepsilon {\cal CS}(\bfh)&=& n \int_0^1 dt\bigg\{\Big( \langle\!\langle
\bfD\varepsilon\ast \bfG_t \ast
\ldots \ast\bfG_t  \rangle\!\rangle+ t \frac d{dt}  \langle\!\langle
\bfD\varepsilon\ast \bfG_t \ast \ldots \ast\bfG_t 
\rangle\!\rangle\Big)\label{deltaCS1}\\
&&-t \langle\!\langle  \bfd \Big(\bfh\ast\bfD\varepsilon\ast \bfG_t \ast \ldots
\ast\bfG_t  +\bfh\ast \bfG_t \ast\bfD\varepsilon\ast \ldots \ast\bfG_t  
+\ldots\0\\
&& \ldots + \bfh\ast \bfG_t \ast \ldots \ast\bfG_t 
\ast\bfD\varepsilon\Big)\rangle\!\rangle\bigg\}\0
\ee
For a proof see Appendix \ref{s:proofs}. 
Upon integrating by parts $\frac d{dt}$, the RHS of the first line becomes
\be
n \langle\!\langle \bfD\varepsilon\ast \bfG\ast \ldots \ast\bfG 
\rangle\!\rangle=
n \langle\!\langle \bfD\left(\varepsilon\ast \bfG\ast \ldots \ast\bfG\right) 
\rangle\!\rangle=
n \langle\!\langle \bfd\left(\varepsilon\ast \bfG\ast \ldots \ast\bfG\right) 
\rangle\!\rangle\label{deltaCS2}
\ee
using the Bianchi identity and \eqref{comm0}. As a consequence
\be
\delta_\varepsilon {\cal CS}(\bfh)&=&n  \langle\!\langle \bfd\bigg(
\varepsilon\ast
\bfG\ast \ldots \ast\bfG- \int_0^1dt\, t\Big(
\bfh\ast\bfD\varepsilon\ast \bfG_t \ast \ldots \ast\bfG_t  +\bfh\ast \bfG_t
\ast\bfD\varepsilon\ast \ldots \ast\bfG_t + \0\\
&& \ldots \ldots+ \bfh\ast \bfG_t \ast \ldots \ast\bfG_t 
\ast\bfD\varepsilon\Big)\bigg)\rangle\!\rangle=0\label{deltaCS3}
\ee
This proves the HS gauge invariance of ${\cal CS}(\bfh)$ in a space of odd
dimension
$d=2n-1$.

\subsection{HS consistent anomalies}

Eq.\eqref{deltaCS3} gives us the form of the primitive
generating
functional for consistent
anomaly in $d=2n-2$ dimension:
\be
{\cal A}(\bfh,\varepsilon)&=& n \langle\!\langle  \varepsilon\ast \bfG\ast
\ldots \ast\bfG- \int_0^1dt \, t\Big(
\bfh\ast\bfD\varepsilon\ast \bfG_t \ast \ldots \ast\bfG_t  +\bfh\ast \bfG_t
\ast\bfD\varepsilon\ast \ldots \ast\bfG_t   +\0\\
&& \ldots \ldots+ \bfh\ast \bfG_t \ast \ldots \ast\bfG_t 
\ast\bfD\varepsilon\Big) \rangle\!\rangle\label{anom1}
\ee
One can write this in a more compact form (see Appendix B):
\be
{\cal A}(\bfh,\varepsilon)&=&n \int_0^1 dt \, (1-t)  \langle\!\langle 
\bfh\ast\bfd\varepsilon\ast \bfG_t \ast \ldots \ast\bfG_t  +\bfh\ast \bfG_t
\ast\bfd\varepsilon\ast \ldots \ast\bfG_t   +\0\\
&& \ldots \ldots+ \bfh\ast \bfG_t \ast \ldots \ast\bfG_t  \ast\bfd\varepsilon
\rangle\!\rangle
\label{anom2}
\ee
where the integrals are over a $4n-4$ dimensional phase space.

Of course we have to prove that the ${\cal A}$ is consistent. 
The shortest way is to promote $\varepsilon$ to anticommuting parameter 
(denoting it with the same symbol). So 
\be
s \bfh = \bfd \varepsilon -i [\bfh\stackrel {\ast}{,} \varepsilon],\quad\quad 
s \varepsilon = \frac i2 [\varepsilon \stackrel {\ast}{,} \varepsilon]=i\, 
\varepsilon\ast  
\varepsilon, \quad\quad {\rm etc.}\label{brsth}
\ee
Using these BRST transform one can prove (see Appendix C)  that
\be 
s  {\cal A}(\bfh,\varepsilon)=0\label{cc}
\ee

With the familiar manipulations one can rewrite \eqref{anom2} as follows
\be
{\cal A}(\bfh,\varepsilon)&=&n \int_0^1 dt \langle\!\langle \varepsilon \ast
\bfG_t\ast \ldots
\ast \bfG_t\,  +i{(1-t)}\varepsilon\ast [
\bfh_t \stackrel{\ast}{,} \bfh \ast \bfG_t \ast \ldots \ast\bfG_t  + \bfG_t
\ast\bfh \ast \ldots \ast\bfG_t   +\0\\
&& \ldots \ldots+ \bfG_t \ast \ldots \ast\bfG_t  \ast\bfh]\rangle\!\rangle
\label{anom3}
\ee
The (unintegrated) anomaly is obtained by differentiating with respect to
$\varepsilon$:
\be
{\cal A}(\bfh)&=&n \int_0^1 dt  \Big{(} \bfG_t\ast \ldots
\ast \bfG_t\,  +i{(1-t)}  [
\bfh_t \stackrel{\ast}{,} \bfh \ast \bfG_t \ast \ldots \ast\bfG_t  + \bfG_t
\ast\bfh \ast \ldots \ast\bfG_t   +\0\\
&& \ldots \ldots+ \bfG_t \ast \ldots \ast\bfG_t 
\ast\bfh]\rangle\!\rangle\label{anom4}
\ee  
which is a $2n-2=d$ form.

This should be understood as the violation the off shell conservation
 ${\cal D}_x^{\ast\mu} J_\mu(x,u)=0$, i.e.
\be
{\cal D}_x^{\ast\mu} J_\mu(x,u)&\sim& \int_0^1 dt  \star\Big{(} \bfG_t\ast
\ldots
\ast \bfG_t\,  +i {(1-t)} [
\bfh_t \stackrel{\ast}{,} \bfh \ast \bfG_t \ast \ldots \ast\bfG_t  + \bfG_t
\ast\bfh \ast \ldots \ast\bfG_t   +\0\\
&& \ldots \ldots+ \bfG_t \ast \ldots \ast\bfG_t  \ast\bfh] \Big{)}
\label{anomconserv}
\ee
where $\star$ denotes the Hodge dual. 

Of course the next problem is to understand if and
when these anomalies appear. This is not difficult. If the matter fermion model
of 
section 2 is constructed with massless chiral fermions (instead of massive Dirac
fermions) it may be 
anomalous and with the same anomaly coefficient as in the ordinary case. In
fact the lowest
order term ${\cal O}(u^0)$ starts with exactly the same term, but consistency
requires 
an infinite tail of anomalous terms. Moreover \eqref{anomconserv} implies that
there are infinite
many conservation laws that are violated, each one with its infinite long
anomaly.

Another question is: do these anomalies exhaust the set of possible anomalies in
HS theories. 
We do not know the answer to this question. We know that in ordinary gauge
theories the 
analogous anomalies are the only consistent ones. But the analogy does  
not seem to be enough to exclude other possible consistent anomalies in HS
theories.

\section{Conclusions}

In this paper we have obtained a series of results on HS effective actions
derived by integrating out 
simple (scalar and fermion) matter fields coupled to external HS sources. In
particular we have developed 
methods to compute current correlators, which are the building blocks of the
effective action. We have shown
how to compute the latter in different ways, either by using the perturbative
formalism introduced in
\cite{BCDGPS} or by means of the more traditional Feynman diagrams. The second
part of the paper 
has dealt with {interpretations of HS objects, their relations and further
developments based 
on the HS symmetry. In particular we have found a precise relation between the
master field $h_a(x,u)$ and 
$h(x,u)$ which couple linearly to fermionic and bosonic matter, respectively.}
The $h_a(x,u)$ potential, with 
its gauge transformations, lends itself to a simple interpretation as HS gauge
field (although the transform contains not only ordinary gauge transformations
but also diffeomorphisms, beside other transformations 
involving all spins). In the third part of the paper we have used this analogy
to compute possible obstructions 
in constructing (even dimensional) HS theories. Such obstructions are the analog
of consistent anomalies in ordinary gauge theories, but, beside the analogy,
there are also significant differences. In particular we have seen that,
whenever one such an obstruction appears, there are infinite many anomalous
currents, and each anomaly consists of an infinite expressions in the local
fields. In passing we have shown that one can construct odd dimensional CS
actions.

The double nature of $h_a(x,u)$ will be at the root of the next two papers, II
and III. As a gauge-like object 
it will be used in paper II to construct HS YM-like theories. As a
{geometry-like} object it will allow us in paper III to interpret the HS
geometry within a scheme analogous to teleparallelism.

\vskip 1cm
{\bf Acknowledgements.}
 
This research has been supported 
by the University of
Rijeka under the research support No.~13.12.1.4.05.
The research of S.G. has been supported by the Israel Science Foundation (ISF),
grant No. 244/17.

\appendix
\section*{Appendices}

\section{Derivation of \eqref{WI2FT} and \eqref{WI2FT'}}
\label{s:WI2FT}

The Fourier transform of the first line of \eqref{WI2} is elementary. Let us see
next
\be
&& \left[\delta(z-y)\delta(t-v)\stackrel{\ast}{,}
\EW^{(2)}_{\nu\mu}(y,v;x,u)\right]\label{ddW2}\\
&=& {i}\int \frac{d^d r}{(2\pi)^d} \int d^d w\,\Bigg( e^{-ir\cdot(z-y)}
e^{iw\cdot(t-v)} 
e^{\frac i2 \left( \stackrel {\leftarrow}{\partial_y}\cdot \stackrel
{\rightarrow}{\partial_v}-
 \stackrel {\rightarrow}{\partial_y}\cdot \stackrel
{\leftarrow}{\partial_v}\right)}\big\langle
J_\nu(y,v) J_\mu(x,u)\big\rangle\0\\
&&- \big\langle J_\nu(y,v) J_\mu(x,u)\big\rangle 
 e^{\frac i2 \left( \stackrel {\leftarrow}{\partial_y}\cdot \stackrel
{\rightarrow}{\partial_v}-
 \stackrel {\rightarrow}{\partial_y}\cdot \stackrel
{\leftarrow}{\partial_v}\right)} 
 e^{-ir\cdot(z-y)} e^{iw\cdot(t-v)} \Bigg)\0\\
&=& {i} \frac{d^d r}{(2\pi)^d}   \int d^d w \, e^{-ir\cdot(z-y)}
e^{iw\cdot(t-v)}
\Bigg(
 e^{-\frac 12 \left(r\cdot \stackrel {\rightarrow}{\partial_v}+ w \cdot
 \stackrel {\rightarrow}{\partial_y}\right)} \big\langle
J_\nu(y,v) J_\mu(x,u)\big\rangle\0\\
&&- \big\langle J_\nu(y,v) J_\mu(x,u)\big\rangle e^{\frac 12 \left(r\cdot
\stackrel {\leftarrow}{\partial_v}+ w \cdot
 \stackrel {\leftarrow}{\partial_y}\right)} \Bigg{)}\0\\
&=& {i}\frac{d^d r}{(2\pi)^d}   \int d^d w\,  e^{-ir\cdot(z-y)} e^{iw\cdot(t-v)}
\Big(
\big\langle
J_\nu\left(y-\frac w2,v-\frac r2\right) J_\mu(x,u)\big\rangle\0\\
&&-
\big\langle
J_\nu\left(y+\frac w2,v+\frac r2\right) J_\mu(x,u)\big\rangle\Big)\0
\ee
The Fourier transform of this is
\be
&&\int d^dx\, e^{ik_1\cdot x} \int dy \, e^{ik_2\cdot y} \int d^dz \,
e^{-iq\cdot z} 
\left[\delta(z-y)\delta(t-v)\stackrel{\ast}{,}
\EW^{(2)}_{\nu\mu}(y,v;x,u)\right]\label{FTddW2}\\
&=& {i}\int d^dx\, e^{ik_1\cdot x} \int dy \, e^{ik_2\cdot y} \int d^dz \,
 e^{-iq\cdot z} \int d^d r \int d^d w\, 
 e^{-ir\cdot(z-y)} e^{iw\cdot(t-v)}\0\\
&&\times \,\Big(
\big\langle
J_\nu\left(y-\frac w2,v-\frac r2\right) J_\mu(x,u)\big\rangle -\big\langle
J_\nu\left(y+\frac w2,v+\frac r2\right) J_\mu(x,u)\big\rangle\Big)\0 
\ee
Integrating over $z$ we find $\delta(q+r)$, and over $w$ 
we find $\delta \left(t-v-\frac {k_1}2\right)$ for the first piece and  
$\delta \left(t-v+\frac {k_1}2\right)$ 
for the second. Next, performing the Fourier transform, one finds
\be
&&\int d^dx\, e^{ik_1\cdot x} \int dy \, e^{ik_2\cdot y} \int d^dz \,
e^{-iq\cdot z} 
\left[\delta(z-y)\delta(t-v)\stackrel{\ast}{,}
\EW^{(2)}_{\nu\mu}(y,v;x,u)\right]\label{FTddW2'}\\
&=&{i}\delta\left(t-v-\frac {k_1}2\right) 
\big\langle \widetilde J_\mu(k_1,u) \widetilde J_\nu\left(-k_1,v+\frac q2\right)
\big\rangle\0\\
&&-{i} \delta\left(t-v+\frac {k_1}2\right) 
\big\langle \widetilde J_\mu(k_1,u) \widetilde J_\nu\left(-k_1,v-\frac q2\right)
\big\rangle\0
\ee

\section{Explicit calculation of 0- and 1-point function}
\label{s:0and1pt}

\subsection{ The first term of (\ref{EW})}
\label{ss:0pt}

Let us  calculate the vacuum energy for fermions.  First, the determinant of the
inverse propagator can be written as
\be 
det(-\widehat{\slashed{P}}-m)=det(-\widehat{P}^2+m^2)^{1/2}
\ee 
The $\EW^{(0)}[h]$ part of the effective action is
\be 
\EW^{(0)}[h]&=&\frac N2 ln[det(-\widehat{P}^2+m^2)]\0\\
&=& \frac N2 \Tr[ln(-\widehat{P}^2+m^2)]\0\\
&=& \frac N2 \int\frac{d^du}{(2\pi)^d}\tr \left(\langle u
|ln(-\widehat{P}^2+m^2)| u
\rangle\right)\0\\
&=& \frac N2 \int\frac{d^du}{(2\pi)^d}ln(-u^2+m^2)\tr \left(\langle u |
u\rangle\right)\0\\
&=& 2^{\lfloor\frac d2 \rfloor-1} \, N \int\frac{d^du}{(2\pi)^d}
ln(-u^2+m^2)\int d^dx
\ee 
where we have assumed the symbol of $\widehat P$ is $u$.
To calculate the integral we  make a Wick rotation $u^0\rightarrow iu^0_E$:
\be 
\EW^{(0)}[h]
&=& 2^{\lfloor\frac d2 \rfloor-1} \, N\,i \int\frac{d^du_E}{(2\pi)^d}
ln(u_E^2+m^2)\int d^dx\0\\
&=& - 2^{\lfloor\frac d2 \rfloor-1-d} \, N\,i\, m^d\pi^{-d/2}\Gamma\left(-\frac
d2\right)\int d^dx
\ee
This is infinite. Since we are not interested in this paper in cosmological
constant problems, we will set it to zero.
 
\subsection{The second term of (\ref{EW})}

Let us now focus on the tadpole contribution.
The $\EW^{(1)}[h]$ part of the effective action is
\be 
\EW^{(1)}[h]&=&\int d^dx\int\frac{d^du}{(2\pi)^d}\EW^{(1)}_{a}(x,u)h^a(x,u)
\ee 
where $\EW^{(1)}_{a}(x,u)$ is given by (\ref{EW1}) and $h^a(x,u)$ is defined
with (\ref{hmmm}). Note that the integrals in $\EW^{(1)}[h]$ over the odd number
of
$u$'s vanish. We are left with:
\be 
\EW^{(1)}[h]=- 2^{\lfloor\frac d2 \rfloor}\, N\,i\,\int
d^dx\int\frac{d^du}{(2\pi)^d}\int_{-\infty}^{\infty} \frac{d\omega}{2\pi }\,
\frac{e^{i\omega \epsilon}}{\omega}\sum_{n=0}^{\infty}\frac 1{(2n+1)!}
h_{(2n+2)}^{a a_1\ldots a_{2n+1}} (x)\, \frac{u_a u_{a_1}\ldots
u_{a_{2n+1}}}{u^2-m^2}
\ee
The integral over the $u$ momenta is
\be 
\int\frac{d^du}{(2\pi)^d}\, \frac{u_a u_{a_1}\ldots
u_{a_{2n+1}}}{u^2-m'^2}&=&\frac{(2n+1)!!}{\prod_{i=0}^n
(d+2i)}\eta_{(a a_1}\ldots\eta_{\mu_{2n}a_{2n+1})}\int\frac{d^du}{(2\pi)^d}
\, \frac{u^{2n+2}}{u^2-m^2}\\
&=& \,
i\,\frac{(-1)^n(2n+1)!!}{2^{d+n+1}}\pi^{-d/2}m'^{2n+d}\Gamma\left(-n-\frac
d2\right)\eta_{(a a_1}\ldots\eta_{\mu_{2n}a_{2n+1})}\0
\ee
Contracting $\eta$'s  with  $h_{(2n+2)}^{\mu\mu_1\ldots\mu_{2n+1}}$ gives the
$(n+1)$st trace of the spin $(2n+2)$ field, which we will denote by
$h_{(2n+2)}^{[n+1]} $. Altogether we have
\be 
\EW^{(1)}[h]&=& 2^{\lfloor\frac d2 \rfloor-d-1}\, N\,\pi^{-d/2}\int
d^dx\int_{-\infty}^{\infty} \frac{d\omega}{2\pi }\, \frac{e^{i\omega
\epsilon}}{\omega}\sum_{n=0}^{\infty}\frac {(-1)^n(2n+1)!!}{2^n(2n+1)!}
\Gamma\left(-n-\frac d2\right) \0\\
&\times&\!(m-i\omega-\varepsilon)^{2n+d} h_{(2n+2)}^{[n+1]} (x)\0\\
&=& 2^{\lfloor\frac d2 \rfloor-d-2}\,i\, N\,\pi^{-d/2}\int
d^dx\sum_{n=0}^{\infty}\frac {(-1)^n(2n+1)!!}{2^n(2n+1)!} \Gamma\left(-n-\frac
d2\right)  m^{2n+d} h_{(2n+2)}^{[n+1]} (x)
\ee

Finally, the effective action up to first order in $h$ is 
\be 
\EW[h]
&\approx& - 2^{\lfloor\frac d2 \rfloor-1-d} \, N\,i\,
m^d\pi^{-d/2}\Gamma\left(-\frac d2\right)\int d^dx\left(1-\frac 12 h_{(2)}^{[1]}
(x)\right)\0\\
&+& 
2^{\lfloor\frac d2 \rfloor-d-2}\,i\, N\,\pi^{-d/2}\int
d^dx\sum_{n=1}^{\infty}\frac {(-1)^n(2n+1)!!\Gamma\left(-n-\frac
d2\right)}{2^n(2n+1)!}   m^{2n+d} h_{(2n+2)}^{[n+1]} (x)+\ldots\label{EWh1}
\ee
One would expect that the first line in (\ref{EWh1})  corresponds to the
cosmological constant term which is of the form
\be 
\Lambda\int d^dx\sqrt{g}\sim\Lambda\int d^dx\left(1+\frac 12 h_{(2)}^{[1]}
(x)+\ldots\right)
\ee
{ We notice that there is a minus sign next to $ h_{(2)}^{[1]} (x)$ term  in the
first
line of (\ref{EWh1}), instead of plus. On the other hand this result is infinite
for even $d$, 
and therefore scheme dependent. Since in this paper we are interested in  flat
spacetime we 
will assume that the first line of (\ref{EWh1}) can be renormalized to 0.}

\section{Gauge transformations in the bosonic model}
\label{s:gaugetransf-bosonic}

 {In this appendix we present the general form of the gauge
transformations in the bosonic model. In order to keep the formulae as simple as
possible, for the time being we define the $\ast$-product as
\be
\alpha(x,u)\ast \beta(x,u)=\alpha(x,u)e^{{i}(\overleftarrow\partial_x \cdot
\overrightarrow\partial_u-\overrightarrow\partial_x \cdot
\overleftarrow\partial_u)} \beta(x,u)\,.
\ee}

We consider first the general case
\begin{eqnarray*}
h\left(x,u\right) & = &
\sum_{p=0}^{s'}\frac{1}{p!}\overset{\left(p\right)}{h}^{\mu_{1}\ldots\mu_{p}}u_{
\mu_{1}}\ldots u_{\mu_{p}}\\
{\eta}(x,u) & = &
\frac{1}{\left(s-1\right)!}\overset{\left(s-1\right)}{\epsilon}^{\mu_{1}
\ldots\mu_{s-1}}u_{\mu_{1}}\ldots u_{\mu_{s-1}}.
\end{eqnarray*}

{
\begin{eqnarray*}
\delta_{\epsilon}^{\left(1\right)}h\left(x,u\right) & = &
-\frac{i}{2}\left[h\left(x,u\right)\stackrel{\ast}{,}\epsilon\left(x,
u\right)\right]\\
 & = &
h\left(x,u\right)\sin\left(\overleftarrow{\partial}_{x}
\cdot\overrightarrow{\partial}_{u}-\overrightarrow{\partial}_{x}
\cdot\overleftarrow{\partial}_{u}\right)\epsilon\left(x,u\right)\\
 & = &
\sum_{r=0}^{\left[\frac{s+s'-2}{2}\right]}\frac{\left(-\right)^{r}}{
\left(2r+1\right)!}h\left(x,
u\right)\left(\overleftarrow{\partial}_{x}\cdot\overrightarrow{\partial}_{u}
-\overrightarrow{\partial}_{x}\cdot\overleftarrow{\partial}_{u}\right)^{2r+1}
\epsilon\left(x,u\right)\\
 & = &
\sum_{r=0}^{\left[\frac{s+s'-2}{2}\right]}\sum_{k=0}^{2r+1}\frac{
\left(-\right)^{r-k+1}}{\left(2r+1\right)!}\binom{
2r+1}{k}h\left(x,u\right)\left(\overleftarrow{\partial}_{x}\cdot\overrightarrow{
\partial}_{u}\right)^{k}\left(\overrightarrow{\partial}_{x}\cdot\overleftarrow{
\partial}_{u}\right)^{2r+1-k}\epsilon\left(x,u\right)\\
 & = &
\sum_{r=0}^{\left[\frac{s'-1}{2}\right]}\frac{\left(-\right)^{r+1}}{
\left(2r+1\right)!}h\left(x,
u\right)\left(\overrightarrow{\partial}_{x}\cdot\overleftarrow{\partial}_{u}
\right)^{2r+1}\epsilon\left(x,u\right)\\
 &  &
+\sum_{k=1}^{s-1}\sum_{r=\left[\frac{k}{2}\right]}^{\left[\frac{
k-1+s'}{2}\right]}\frac{\left(-\right)^{r-k+1}}{\left(2r+1\right)!}\binom{2r+1}{
k}h\left(x,u\right)\left(\overleftarrow{\partial}_{
x}\cdot\overrightarrow{\partial}_{u}\right)^{k}\left(\overrightarrow{\partial}_{
x}\cdot\overleftarrow{\partial}_{u}\right)^{2r+1-k}\epsilon\left(x,u\right)\\
 & = &
\sum_{k=0}^{s-1}\sum_{r=\left[\frac{k}{2}\right]}^{\left[\frac{k-1+s'
}{2}\right]}\frac{\left(-\right)^{r-k+1}}{\left(2r+1\right)!}\binom{2r+1}{k}
h\left(x,u\right)\left(\overleftarrow{\partial}_{x}
\cdot\overrightarrow{\partial}_{u}\right)^{k}\left(\overrightarrow{\partial}_{x}
\cdot\overleftarrow{\partial}_{u}\right)^{2r+1-k}\epsilon\left(x,u\right),
\end{eqnarray*}
}

The condition $0\leq r\leq\left[\frac{s'-1}{2}\right]$ for the first
term comes from the requirement $k=0$ and $2r+1\leq s'$. The bounds
on $k$ in the second term come from $k\leq s-1$ . The bounds on
$r$ come from the condition $k\leq2r+1$, i.e. $r\geq\left[\frac{k}{2}\right]$
and $2r+1-k\leq s'$,i.e. $r\leq\left[\frac{k-1+s'}{2}\right]$ . 

{
\begin{multline*}
\left.\frac{\partial}{\partial u_{\mu_{1}}}\ldots\frac{\partial}{\partial
u_{\mu_{t}}}\sum_{k=0}^{s-1}\sum_{r=\left[\frac{k}{2}\right]}^{\left[
\frac{k-1+s'}{2}\right]}\frac{\left(-\right)^{r-k+1}}{\left(2r+1\right)!}
\binom{2r+1}{k}h\left(x,
u\right)\left(\overleftarrow{\partial}_{x}\cdot\overrightarrow{\partial}_{u}
\right)^{k}\left(\overrightarrow{\partial}_{x}\cdot\overleftarrow{\partial}_{u}
\right)^{2r+1-k}\epsilon\left(x,u\right)\right|_{u=0}=\\
=\sum_{k=0}^{s-1}\sum_{r=\left[\frac{k}{2}\right]}^{\left[\frac{
k-1+s'}{2}\right]}\frac{\left(-\right)^{r-k+1}}{\left(2r+1\right)!}\binom{2r+1}{
k}\sum_{i=0}^{t}\binom{t}{t-i}\\
\left.\frac{\partial}{\partial
u_{\left(\mu_{i+1}\right.}}\ldots\frac{\partial}{\partial
u_{\mu_{t}}}h\left(x,u\right)\left(\overleftarrow{\partial}_{x}
\cdot\overrightarrow{\partial}_{u}\right)^{k}\left(\overrightarrow{\partial}_{x}
\cdot\overleftarrow{\partial}_{u}\right)^{2r+1-k}\frac{\partial}{\partial
u_{\mu_{1}}}\ldots\frac{\partial}{\partial
u_{\left.\mu_{i}\right)}}\epsilon\left(x,u\right)\right|_{u=0}\\
=\sum_{k=0}^{s-1}\sum_{r=\left[\frac{k}{2}\right]}^{\left[\frac{
k-1+s'}{2}\right]}\frac{\left(-\right)^{r-k+1}}{\left(2r+1\right)!}\binom{2r+1}{
k}\sum_{i=0}^{t}\binom{t}{t-i}\\
\partial_{\nu_{1}}\ldots\partial_{\nu_{k}}\overset{\left(t+2r+1-k-i\right)}{h}^{
\nu_{k+1}\ldots\nu_{2r+1}\left(\mu_{i+1}\ldots\mu_{t}\right.}
\left(x\right)\partial_{\nu_{k+1}}\ldots\partial_{\nu_{2r+1}}\stackrel{
\left(k+i\right)}{\epsilon}^{\left|\nu_{1}\ldots\nu_{k}\right|\left.\mu_{1}
\ldots\mu_{i}\right)}\left(x\right)\\
\end{multline*}
}

This implies the constraints $i+k=s-1$ and $t+2r+1-k-i\leq s'$, implying
$t\leq s+s'-2$. Therefore $0\leq i=s-k-1\leq t$ which means $s-t-1\leq k\leq
s-1$.
If $t\leq s-2$ , this is the most restrictive condition; if $s-1\leq t\leq
s+s'-2$
then $0\leq k\leq s-1$. On the other hand we have $2r+1\geq k\geq s-t-1$,i.e.
$r\geq\left[\frac{k}{2}\right]$ and $t+2r+1-k-i\leq s$,i.e. $2r+1\leq
s+s'-t-1\leq s'+k$.

So, for $t\leq s-2$ ,

{

\begin{multline*}
\sum_{k=s-t-1}^{s-1}\sum_{r=\left[\frac{k}{2}\right]}^{\left[\frac{
s+s'-t-2}{2}\right]}\frac{\left(-\right)^{r-k+1}}{\left(2r+1\right)!}\binom{2r+1
}{k}\binom{t}{s-k-1}\\
\partial_{\nu_{1}}\ldots\partial_{\nu_{k}}\overset{\left(t-s+1+2r+1\right)}{h}^{
\nu_{k+1}\ldots\nu_{2r+1}\left(\mu_{s}\ldots\mu_{t+k}\right.}
\left(x\right)\partial_{\nu_{k+1}}\ldots\partial_{\nu_{2r+1}}\stackrel{
\left(s-1\right)}{\epsilon}^{\left|\nu_{1}\ldots\nu_{k}\right|\left.\mu_{k+1}
\ldots\mu_{s-1}\right)}\left(x\right)
\end{multline*}
}

For $s-1\leq t\leq s+s'-2$,

{
\begin{multline*}
\sum_{k=0}^{s-1}\sum_{r=\left[\frac{k}{2}\right]}^{\left[\frac{
s+s'-t-2}{2}\right]}\frac{\left(-\right)^{r-k+1}}{\left(2r+1\right)!}\binom{2r+1
}{k}\binom{t}{s-k-1}\\
\partial_{\nu_{1}}\ldots\partial_{\nu_{k}}\overset{\left(t-s+1+2r+1\right)}{h}^{
\nu_{k+1}\ldots\nu_{2r+1}\left(\mu_{s}\ldots\mu_{t+k}\right.}
\left(x\right)\partial_{\nu_{k+1}}\ldots\partial_{\nu_{2r+1}}\stackrel{
\left(s-1\right)}{\epsilon}^{\left|\nu_{1}\ldots\nu_{k}\right|\left.\mu_{k+1}
\ldots\mu_{s-1}\right)}\left(x\right)
\end{multline*}
}

All $\delta\overset{\left(t\right)}{h}^{\mu_{1}\ldots\mu_{t}}$ with
$t\geq s+s'-1$ are vanishing.

\subsection{Truncated transformations}

Let us work out some examples explicitly. For $s=1$ and $s'=3$ we
get 
{
\begin{eqnarray*}
\delta\overset{\left(0\right)}{h} & = &
\delta^{\left(1\right)}\overset{\left(0\right)}{h}\\
 & = &
-\overset{\left(1\right)}{h}^{\nu_{1}}\partial_{\nu_{1}}
\epsilon+\frac{1}{6}\overset{\left(3\right)}{h}^{\nu_{1}\nu_{2}\nu_{3}}
\partial_{\nu_{1}}\partial_{\nu_{2}}\partial_{\nu_{3}}\epsilon\\
\delta\overset{\left(1\right)}{h}^{\mu_{1}} & = &
\delta^{\left(0\right)}\overset{\left(1\right)}{h}^{\mu_{1}}+\delta^{
\left(1\right)}\overset{\left(1\right)}{h}^{\mu_{1}}\\
 & = &
\partial^{\mu_{1}}\epsilon-\overset{\left(2\right)}{h}^{\nu_{1}\mu_{1
}}\partial_{\nu_{1}}\epsilon\\
\delta\overset{\left(2\right)}{h}^{\mu_{1}\mu_{2}} & = &
\delta^{\left(1\right)}\overset{\left(1\right)}{h}^{\mu_{1}\mu_{2}}\\
 & = &-
\overset{\left(3\right)}{h}^{\nu_{1}\mu_{1}\mu_{2}}\partial_{\nu_{1}}
\epsilon\\
\delta\overset{\left(3\right)}{h}^{\mu_{1}\mu_{2}\mu_{3}} & = & 0\\
\vdots & \vdots & \vdots
\end{eqnarray*}
}

For $s=2$ and $s'=3$ we get
{
\begin{eqnarray*}
\delta\overset{\left(0\right)}{h} & = &
\delta^{\left(1\right)}\overset{\left(0\right)}{h}\\
 & = &
\epsilon^{\nu_{1}}\partial_{\nu_{1}}\overset{\left(0\right)}{h}
-\frac{1}{2}\partial_{\nu_{1}}\overset{
\left(2\right)}{h}^{\nu_{2}\nu_{3}}\partial_{\nu_{2}}\partial_{\nu_{3}}\epsilon^
{\nu_{1}}\\
\delta\overset{\left(1\right)}{h}^{\mu_{1}} & = &
\delta^{\left(1\right)}\overset{\left(1\right)}{h}^{\mu_{1}}\\
 & = &
\epsilon^{\nu_{1}}\partial_{\nu_{1}}\overset{\left(1\right)}{h}
^{\mu_{1}}-\overset{\left(1\right)}{h}^{\nu_{1}}\partial_{\nu_{1}}\epsilon^{\mu_
{1}}\\
 &  &
+\frac{1}{6}\overset{\left(3\right)}{h}^{\nu_{
1}\nu_{2}\nu_{3}}\partial_{\nu_{1}}\partial_{\nu_{2}}\partial_{\nu_{3}}\epsilon^
{\mu_{1}}-\frac{1}{2}\partial_{\nu_{1}}\overset{
\left(3\right)}{h}^{\nu_{2}\nu_{3}\mu_{1}}\partial_{\nu_{2}}\partial_{\nu_{3}}
\epsilon^{\nu_{1}}\\
\delta\overset{\left(2\right)}{h}^{\mu_{1}\mu_{2}} & = &
\delta^{\left(0\right)}\overset{\left(2\right)}{h}^{\mu_{1}\mu_{2}}+\delta^{
\left(1\right)}\overset{\left(2\right)}{h}^{\mu_{1}\mu_{2}}\\
 & = & 2\partial^{\left(\mu_{1}\right|}\epsilon^{\left.\mu_{2}\right)}\\
 &  &
+\epsilon^{\nu_{1}}\partial_{\nu_{1}}\overset{\left(2\right)}{h
}^{\mu_{1}\mu_{2}}-2\overset{\left(2\right)}{h}^{\nu_{1}\left(\mu_{1}\right.}
\partial_{\nu_{1}}\epsilon^{\left.\mu_{2}\right)}\\
\delta\overset{\left(3\right)}{h}^{\mu_{1}\mu_{2}\mu_{3}} & = &
\epsilon^{\nu_{1}}\partial_{\nu_{1}}\overset{\left(3\right)}{h}
^{\mu_{1}\mu_{2}\mu_{3}}-3\overset{\left(3\right)}{h}^{\nu_{1}\left(\mu_{1}\mu_{
2}\right.}\partial_{\nu_{1}}\epsilon^{\left.\mu_{3}\right)}\\
\delta\overset{\left(4\right)}{h}^{\mu_{1}\mu_{2}\mu_{3}\mu_{4}} & = & 0\\
\vdots & \vdots & \vdots
\end{eqnarray*}
}

For $s=3$ and $s'=3$  we get 
{
\begin{eqnarray*}
\delta\overset{\left(0\right)}{h} & = &
\delta^{\left(1\right)}\overset{\left(0\right)}{h}\\
 & = &
\frac{1}{2}\partial_{\nu_{1}}
\partial_{\nu_{2}}\overset{\left(1\right)}{h}^{\nu_{3}}\partial_{\nu_{3}}
\epsilon^{\nu_{1}\nu_{2}}\\
 &  &
-\frac{1}{12}\partial_{\nu_{1}}\partial_{\nu_
{2}}\overset{\left(3\right)}{h}^{\nu_{3}\nu_{4}\nu_{5}}\partial_{\nu_{3}}
\partial_{\nu_{4}}\partial_{5}\epsilon^{\nu_{1}\nu_{2}}\\
\delta\overset{\left(1\right)}{h}^{\mu_{1}} & = &
\delta^{\left(1\right)}\overset{\left(1\right)}{h}^{\mu_{1}}\\
 & = &
\epsilon^{\nu_{1}\mu_{1}}\partial_{\nu_{1}}\overset{
\left(0\right)}{h}\\
 &  &
-\frac{1}{2}\partial_{\nu_{1}}\overset{
\left(2\right)}{h}^{\nu_{2}\nu_{3}}\partial_{\nu_{2}}\partial_{\nu_{3}}\epsilon^
{\nu_{1}\mu_{1}}\\
 &  &
+\frac{1}{2}\partial_{\nu_{1}}\partial_{\nu_{2
}}\overset{\left(2\right)}{h}^{\nu_{3}\mu_{1}}\partial_{\nu_{3}}\epsilon^{\nu_{1
}\nu_{2}}\\
\delta\overset{\left(2\right)}{h}^{\mu_{1}\mu_{2}} & = &
\delta^{\left(1\right)}\overset{\left(2\right)}{h}^{\mu_{1}\mu_{2}}\\
 & = &
2\epsilon^{\nu_{1}\left(\mu_{1}\right.}\partial_{\nu_{1}}
\overset{\left(1\right)}{h}^{\left.\mu_{2}\right)}-\overset{\left(1\right)}{h}^{
\nu_{1}}\partial_{\nu_{1}}\epsilon^{\mu_{1}\mu_{2}}\\
 &  &
+\frac{1}{6}\overset{\left(3\right)}{h}^{\nu_{1}
\nu_{2}\nu_{3}}\partial_{\nu_{1}}\partial_{\nu_{2}}\partial_{\nu_{3}}\epsilon^{
\mu_{1}\mu_{2}}\\
 &  &
-\partial_{\nu_{1}}\overset{\left(3\right)}{h}^{\nu_
{2}\nu_{3}\left(\mu_{1}\right.}\partial_{\nu_{2}}\partial_{\nu_{3}}\epsilon^{
\left|\nu_{1}\right|\left.\mu_{2}\right)}\\
 &  &
+\frac{1}{2}\partial_{\nu_{1}}\partial_{\nu_{2
}}\overset{\left(3\right)}{h}^{\nu_{3}\mu_{1}\mu_{2}}\partial_{\nu_{3}}\epsilon^
{\nu_{1}\nu_{2}}\\
\delta\overset{\left(3\right)}{h}^{\mu_{1}\mu_{2}\mu_{3}} & = &
\delta^{\left(0\right)}\overset{\left(3\right)}{h}^{\mu_{1}\mu_{2}\mu_{3}}
+\delta^{\left(1\right)}\overset{\left(3\right)}{h}^{\mu_{1}\mu_{2}\mu_{3}}\\
 & = & 3\partial^{\left(\mu_{1}\right.}\epsilon^{\left.\mu_{2}\mu_{3}\right)}\\
 &  &
+3\epsilon^{\nu_{1}\left(\mu_{1}\right.}\partial_{\nu_{1}}
\overset{\left(2\right)}{h}^{\left.\mu_{2}\mu_{3}\right)}-3\overset{
\left(2\right)}{h}^{\nu_{1}\left(\mu_{1}\right.}\partial_{\nu_{1}}\epsilon^{
\left.\mu_{2}\mu_{3}\right)}\\
\delta\overset{\left(4\right)}{h}^{\mu_{1}\mu_{2}\mu_{3}\mu_{4}} & = &
\delta^{\left(1\right)}\overset{\left(4\right)}{h}^{\mu_{1}\mu_{2}\mu_{3}\mu_{4}
}\\
 & = &
4\epsilon^{\nu_{1}\left(\mu_{1}\right.}\partial_{\nu_{1}}
\overset{\left(3\right)}{h}^{\left.\mu_{1}\mu_{2}\mu_{3}\right)}-6\overset{
\left(3\right)}{h}^{\nu_{1}\left(\mu_{1}\mu_{2}\right.}\partial_{\nu_{1}}
\epsilon^{\left.\mu_{3}\mu_{4}\right)}\\
\delta\overset{\left(5\right)}{h}^{\mu_{1}\mu_{2}\mu_{3}\mu_{4}\mu_{5}} & = &
0\\
\vdots & \vdots & \vdots
\end{eqnarray*}
}

\section{Some proofs }
\label{s:proofs}

\subsection{Proof of eq.\eqref{deltaCS1}}

\be
\delta_\varepsilon {\cal CS}(\bfh)&=& n\int_0^1 dt\Big( \langle\!\langle
\bfD\varepsilon\ast \bfG_t \ast \ldots \ast\bfG_t  \rangle\!\rangle{+}t
\langle\!\langle  \bfh  \ast  \bfd_t\bfD\varepsilon \ast \ldots \ast\bfG_t 
\rangle\!\rangle{+} \ldots\label{deltaCS1'}\\
&&\dots {+} t \langle\!\langle  \bfh  \ast \bfG_t  \ast \ldots
\ast\bfd_t\bfD\varepsilon \rangle\!\rangle\Big)\0\\
&=& n\int_0^1 dt\Big( \langle\!\langle \bfD\varepsilon\ast \bfG_t \ast \ldots
\ast\bfG_t  \rangle\!\rangle{+}t \langle\!\langle  \bfh  \ast 
\bfd_t(\bfD\varepsilon \ast \ldots \ast\bfG_t ) \rangle\!\rangle+ \ldots\0\\
&&\dots {+} t \langle\!\langle  \bfh  \ast\bfd_t( \bfG_t  \ast \ldots
\ast\bfD\varepsilon )\rangle\!\rangle\Big)\0\\
&=&n \int_0^1 dt\Big( \langle\!\langle \bfD\varepsilon\ast \bfG_t \ast \ldots
\ast\bfG_t  \rangle\!\rangle{+}t \langle\!\langle   \frac {d \bfG_t}{dt} \ast 
\bfD\varepsilon \ast \ldots \ast\bfG_t  \rangle\!\rangle+ \ldots\0\\
&&\dots {+} t \langle\!\langle \frac {d \bfG_t}{dt} \ast \bfG_t  \ast \ldots
\ast\bfD\varepsilon \rangle\!\rangle\Big)\0\\
&&- t\langle\!\langle  \bfd \Big(\bfh\ast\bfD\varepsilon\ast \bfG_t \ast \ldots
\ast\bfG_t  +\bfh\ast \bfG_t \ast\bfD\varepsilon\ast \ldots \ast\bfG_t  
+\ldots\0\\
&& \ldots + \bfh\ast \bfG_t \ast \ldots \ast\bfG_t 
\ast\bfD\varepsilon\Big)\rangle\!\rangle\0\\
&=&n \int_0^1 dt\bigg\{\Big( \langle\!\langle \bfD\varepsilon\ast \bfG_t \ast
\ldots \ast\bfG_t  \rangle\!\rangle+ t \frac d{dt}  \langle\!\langle
\bfD\varepsilon\ast \bfG_t \ast \ldots \ast\bfG_t  \rangle\!\rangle\Big)\0\\
&&-t \langle\!\langle  \bfd \Big(\bfh\ast\bfD\varepsilon\ast \bfG_t \ast \ldots
\ast\bfG_t  +\bfh\ast \bfG_t \ast\bfD\varepsilon\ast \ldots \ast\bfG_t  
+\ldots\0\\
&& \ldots + \bfh\ast \bfG_t \ast \ldots \ast\bfG_t 
\ast\bfD\varepsilon\Big)\rangle\!\rangle\bigg\}\0
\ee

\section{From eq.\eqref{anom1} to eq.\eqref{anom2}}

The starting point is 
\be
{\cal A}(\bfh,\varepsilon)&=& n \langle\!\langle  \varepsilon\ast \bfG\ast
\ldots \ast\bfG- \int_0^1dt \, t\Big(
\bfh\ast\bfD\varepsilon\ast \bfG_t \ast \ldots \ast\bfG_t  +\bfh\ast \bfG_t
\ast\bfD\varepsilon\ast \ldots \ast\bfG_t   +\0\\
&& \ldots \ldots+ \bfh\ast \bfG_t \ast \ldots \ast\bfG_t 
\ast\bfD\varepsilon\Big) \rangle\!\rangle\label{anom1b}
\ee
Let us consider the term in round bracket and decompose 
$\bfD\varepsilon= \bfd \varepsilon -i [\bfh \stackrel{\ast}{,} \varepsilon]$. We
consider first
the second term, i.e.
\be
-it\, \langle\!\langle
\bfh\ast[\bfh \stackrel{\ast}{,} \varepsilon]\ast \bfG_t \ast \ldots \ast\bfG_t 
+\bfh\ast \bfG_t
\ast[\bfh \stackrel{\ast}{,} \varepsilon]\ast \ldots \ast\bfG_t   + \ldots
\ldots+ \bfh\ast \bfG_t \ast \ldots \ast\bfG_t 
\ast[\bfh \stackrel{\ast}{,} \varepsilon]\Big) \rangle\!\rangle\0
\ee
Using \eqref{comm0} this becomes
{\footnotesize
\be
&&-i \langle\!\langle
[ \bfh\stackrel{\ast}{,} \bfh_t] \ast\varepsilon\ast \bfG_t \ast \ldots
\ast\bfG_t  +\bfh\ast \varepsilon
\ast[\bfG_t\stackrel{\ast}{,} \bfh_t]\ast \ldots \ast\bfG_t   + \ldots \ldots
+ \bfh\ast \varepsilon\ast \ldots \ast\bfG_t 
\ast[\bfG_t \stackrel{\ast}{,}\bfh_t]\Big) \rangle\!\rangle\label{anom1b1}\\
&&-i \langle\!\langle
[ \bfh\stackrel{\ast}{,} \bfh_t] \ast \bfG_t\ast\varepsilon \ldots \ast\bfG_t  
+\bfh\ast[\bfG_t\stackrel{\ast}{,} \bfh_t]\ast \varepsilon
\ast \ldots \ast\bfG_t   + \ldots \ldots
+ \bfh\ast\bfG_t\ast \varepsilon\ast \ldots \ast\bfG_t 
\ast[\bfG_t \stackrel{\ast}{,}\bfh_t]\Big) \rangle\!\rangle\0\\
&&\ldots\ldots\0\\
&&-i \langle\!\langle
[ \bfh\stackrel{\ast}{,} \bfh_t] \ast \bfG_t\ast \ldots \ast\bfG_t
\ast\varepsilon
+\bfh\ast[\bfG_t\stackrel{\ast}{,} \bfh_t]\ast 
\ast \ldots \ast\bfG_t \ast \varepsilon  + \ldots \ldots
+ \bfh\ast\bfG_t \ast \ldots \ast\bfG_t 
\ast[\bfG_t \stackrel{\ast}{,}\bfh_t]\ast\varepsilon\Big) \rangle\!\rangle\0
\ee
}
Now we use ${-i}[\bfh_t\stackrel{\ast}{,} \bfh]= \frac {d \bfG_t}{dt} -\bfd 
\bfh$ 
and $[\bfh_t\stackrel{\ast}{,}\bfG_t] =-{i}\bfd \bfG_t$ and \eqref{anom1b} 
becomes
{\footnotesize
\be
&&  \langle\!\langle\frac {d \bfG_t}{dt}\ast \Big( \varepsilon\ast \bfG_t \ast
\ldots \ast\bfG_t
 +\bfG_t\ast\varepsilon \ldots \ast\bfG_t+\ldots\ldots 
+ \bfG_t\ast \ldots \ast\bfG_t
\ast\varepsilon\Big)\rangle\!\rangle\label{anom1b2}\\
&&-\langle\!\langle\bfd \bfh\ast \Big( \varepsilon\ast \bfG_t \ast \ldots
\ast\bfG_t
 +\bfG_t\ast\varepsilon \ldots \ast\bfG_t+\ldots\ldots 
+ \bfG_t\ast \ldots \ast\bfG_t \ast\varepsilon\Big)\rangle\!\rangle\0\\
&& + \langle\!\langle \bfh \ast \varepsilon \ast \bfd \Big(\bfG_t \ast \ldots
\ast\bfG_t\Big)+
\bfh \ast \bfd \Big(\bfG_t \ast \check \varepsilon \ast\ldots
\ast\bfG_t\Big)+\ldots\ldots
+\bfh \ast \bfd\Big(\bfG_t \ast \ldots \ast\bfG_t\Big)\ast
\varepsilon\rangle\!\rangle\0
\ee
}
where $\check{}$ over $\varepsilon$ means that $\bfd$ does not act on it. This
is
{\footnotesize
\be 
&&\frac d{dt} \langle\!\langle   \varepsilon\ast \bfG_t \ast \ldots
\ast\bfG_t\rangle\!\rangle\label{anom1b3}\\
&&-  \langle\!\langle  \bfh \ast \bfd \varepsilon\ast\bfG_t \ast \ldots
\ast\bfG_t +
 \bfh\ast\bfG_t \ast \bfd \varepsilon \ast \ldots \ast\bfG_t+ \ldots\ldots
+ \bfh\ast\bfG_t\ldots \ast\bfG_t\ast \bfd \varepsilon \rangle\!\rangle\0\\
&&-  \langle\!\langle  \bfd \Big(  \bfh \ast \varepsilon\ast\bfG_t \ast \ldots
\ast\bfG_t +
 \bfh\ast\bfG_t \ast \varepsilon \ast \ldots \ast\bfG_t+ \ldots\ldots
+ \bfh\ast\bfG_t\ldots \ast\bfG_t\ast \varepsilon\Big) \rangle\!\rangle\0
\ee
}
The last line vanishes. Disregarding it and adding the first two lines (under
$-\int_0^1 dt$) to the remaining piece in \eqref{anom1b}, i.e.
\be
&&  \langle\!\langle  \varepsilon\ast \bfG\ast
\ldots \ast\bfG- \int_0^1dt \, t\Big(
\bfh\ast\bfd\varepsilon\ast \bfG_t \ast \ldots \ast\bfG_t  +\bfh\ast \bfG_t
\ast\bfd\varepsilon\ast \ldots \ast\bfG_t   +\0\\
&& \ldots \ldots+ \bfh\ast \bfG_t \ast \ldots \ast\bfG_t 
\ast\bfd\varepsilon\Big) \rangle\!\rangle\label{anom1b4}
\ee
and integrating by parts in $t$ the first line of \eqref{anom1b3}, we get
\be
{\cal A}(\bfh,\varepsilon)
&=& n \int_0^1 dt (1-t) \langle\!\langle
\bfh\ast\bfd\varepsilon\ast \bfG_t \ast \ldots \ast\bfG_t  +\bfh\ast \bfG_t
\ast\bfd\varepsilon\ast \ldots \ast\bfG_t \label{Ahe}\\
&&\quad\quad\quad\quad  + \ldots \ldots+ \bfh\ast \bfG_t \ast \ldots \ast\bfG_t 
\ast\bfd\varepsilon \rangle\!\rangle\0
\ee

\section{Proof of eq.\eqref{cc}}

We prove eq.\eqref{cc} for the case $n=3$. In that case
\be 
{\cal A}(\bfh,\varepsilon)&=&3 \int_0^1 dt (1-t) \langle\!\langle
\bfh\ast\bfd\varepsilon\ast \bfG_t + \bfh\ast
\bfG_t\ast\bfd\varepsilon\rangle\!\rangle\label{ccn=3a}\\
&=& {+}\frac 12  \langle\!\langle\bfd\varepsilon\ast(\bfd \bfh \ast \bfh + \bfh
\ast \bfd \bfh) \rangle\!\rangle{-} \frac i2  
\langle\!\langle\bfd\varepsilon\ast\bfh\ast\bfh \ast \bfh\rangle\!\rangle\0
\ee
Applying the BRST operator $s$ 
\be 
s {\cal A}(\bfh,\varepsilon)&=&{+}\frac i2  \langle\!\langle\bfd(\varepsilon\ast
\varepsilon)\ast(\bfd \bfh \ast \bfh + \bfh \ast \bfd \bfh) +
\bfd\varepsilon\ast \left([\bfd \bfh \stackrel{\ast}{,} \varepsilon]-[\bfh
\stackrel{\ast}{,}\bfd  \varepsilon]\right)\ast \bfh\label{ccn=3b}\\
&&+\bfd\varepsilon\ast \bfh \ast  \left([\bfd \bfh \stackrel{\ast}{,}
\varepsilon]-[\bfh \stackrel{\ast}{,}\bfd  \varepsilon]\right)\rangle\!\rangle 
{-}
\frac 12 \langle\!\langle\bfd\varepsilon\ast\bfd \bfh \ast ( \bfd \varepsilon -i
[\bfh \stackrel{\ast}{,} \varepsilon])\0\\
&&+ \bfd\varepsilon \ast ( \bfd \varepsilon -i [\bfh \stackrel{\ast}{,}
\varepsilon])\ast\bfd \bfh\rangle\!\rangle{+} \frac 12\langle\!\langle
\bfd(\varepsilon\ast \varepsilon)\ast\bfh\ast\bfh \ast \bfh\rangle\!\rangle\0\\
{+}\frac i2 \langle\!\langle\bfd\varepsilon\!\!\!\!\!&\ast&\!\!\!\!\!  ( \bfd
\varepsilon -i [\bfh \stackrel{\ast}{,} \varepsilon])\ast\bfh\ast\bfh +
\bfd\varepsilon\ast\bfh\ast ( \bfd \varepsilon -i [\bfh \stackrel{\ast}{,}
\varepsilon])\ast \bfh +  \bfd\varepsilon\ast\bfh\ast \bfh\ast( \bfd \varepsilon
-i [\bfh \stackrel{\ast}{,} \varepsilon])\rangle\!\rangle\0
\ee
Expanding the RHS one can see each term cancels a similar term with opposite
sign, except
\be 
 \langle\!\langle\bfd\varepsilon\ast\bfd\varepsilon\ast\bfd \bfh\rangle\!\rangle
= \langle\!\langle\bfd(\varepsilon\ast\bfd\varepsilon\ast\bfd
\bfh)\rangle\!\rangle=0\label{res1}
\ee
which vanishes by integration, and
\be
\langle\!\langle\bfd\varepsilon\ast\bfh
\ast\bfd\varepsilon\ast\bfh\rangle\!\rangle=
- \langle\!\langle\bfd\varepsilon\ast\bfh\ast
\bfd\varepsilon\ast\bfh\rangle\!\rangle\label{res2}
\ee
which vanishes by symmetry.



\begin{thebibliography}{99}


\bibitem{Maldacena}
X.~O.~Camanho, J.~D.~Edelstein, J.~Maldacena and A.~Zhiboedov,
{\it Causality Constraints on Corrections to the Graviton Three-Point
Coupling,}
  JHEP \textbf {1602} (2016) 020
   [arXiv:1407.5597 [hep-th]].

\bibitem{Vasiliev}  M.~A.~Vasiliev,
{\it Consistent equation for interacting gauge fields of all spins in
(3+1)-dimensions,}
  Phys.\ Lett.\ B \textbf {243} (1990) 378;
   {\it Properties of equations of motion of interacting gauge fields of all
spins
in (3+1)-dimensions,}
  Class.\ Quant.\ Grav.\  \textbf {8} (1991) 1387;
  {\it Algebraic aspects of the higher spin problem,}
  Phys.\ Lett.\ B \textbf {257} (1991) 111;
{\it More on equations of motion for interacting massless fields of all
spins in
(3+1)-dimensions,}
  Phys.\ Lett.\ B \textbf {285} (1992) 225.


\bibitem{FS} 
 D.~Francia and A.~Sagnotti,
  {\it On the geometry of higher spin gauge fields,}
  Class.\ Quant.\ Grav.\  {\bf 20} (2003) S473,
   [Comment.\ Phys.\ Math.\ Soc.\ Sci.\ Fenn.\  {\bf 166} (2004) 165],
   [PoS JHW {\bf 2003} (2003) 005], 
  [hep-th/0212185].
  
D.~Francia and A.~Sagnotti,
{\it Free geometric equations for higher spins,}
  Phys.\ Lett.\ B {\bf 543} (2002) 303  
  [hep-th/0207002].
  
\bibitem{BBvD} F.A.Berends, G.J.H. Burgers and H. Van Dam 
{\it On the theoretical problems in constructing interactions involving
higher-spin masslass particles},
Nucl.\ Phys.\ {\bf B260} (1985) 295-322. 

\bibitem{Fronsdal} C.~Fronsdal,
{\it Massless Fields with Integer Spin,} 
  Phys.\ Rev.\ D {\bf 18} (1978) 3624.

J.~Fang and C.~Fronsdal,
{\it Massless Fields with Half Integral Spin,}
  Phys.\ Rev.\ D {\bf 18} (1978) 3630.

 
\bibitem{Bekaert:2009ud}
  X.~Bekaert, E.~Joung and J.~Mourad,
  ``On higher spin interactions with matter,''
  JHEP {\bf 0905} (2009) 126
  [arXiv:0903.3338 [hep-th]].
  
\bibitem{Bekaert:2010ky}
  X.~Bekaert, E.~Joung and J.~Mourad,
  ``Effective action in a higher-spin background,''
  JHEP {\bf 1102} (2011) 048
  [arXiv:1012.2103 [hep-th]].

\bibitem{BCDGPS}
L.~Bonora, M.~Cvitan, P.~Dominis Prester, S.~Giaccari, M.~Pauli\v{s}i\'{c} and
T.~\v{S}temberga,
  {\it Worldline quantization of field theory, effective actions and $L_\infty$
structure},
  JHEP {\bf 1804} (2018) 095
  [arXiv:1802.02968 [hep-th]].
\bibitem{HZ}
  O.~Hohm and B.~Zwiebach,
 {\it $L_{\infty}$ Algebras and Field Theory,}
  Fortsch.\ Phys.\  {\bf 65} (2017) no.3-4,  1700014
  [arXiv:1701.08824 [hep-th]].


\bibitem{Zwiebach}
 M.~R.~Gaberdiel and B.~Zwiebach,
{\it Tensor constructions of open string theories. 1: Foundations,}
  Nucl.\ Phys.\ B {\bf 505} (1997) 569
  [hep-th/9705038].

 B.~Zwiebach,
{\it Oriented open - closed string theory revisited,}
  Annals Phys.\  {\bf 267} (1998) 193
  [hep-th/9705241].

\bibitem{Stasheff}
 H.~Kajiura and J.~Stasheff,
{\it Homotopy algebras inspired by classical open-closed string field theory,}
  Commun.\ Math.\ Phys.\  {\bf 263} (2006) 553
  [math/0410291 [math-qa]].

\bibitem{Lada1}
  T.~Lada and J.~Stasheff,
{\it Introduction to SH Lie algebras for physicists,}
  Int.\ J.\ Theor.\ Phys.\  {\bf 32} (1993) 1087
  [hep-th/9209099].

\bibitem{Lada2}
T.~Lada and M.~Markl,
{\it Strongly homotopy Lie algebras,}
  [hep-th/9406095].

\bibitem{Barnich}
  G.~Barnich, R.~Fulp, T.~Lada and J.~Stasheff,
{\it The sh Lie structure of Poisson brackets in field theory,}
  Commun.\ Math.\ Phys.\  {\bf 191} (1998) 585
  [hep-th/9702176].

A.~M.~Zeitlin,
{\it String field theory-inspired algebraic structures in gauge theories,}
  J.\ Math.\ Phys.\  {\bf 50} (2009) 063501
  [arXiv:0711.3843 [hep-th]].

 A.~M.~Zeitlin,
{\it Conformal Field Theory and Algebraic Structure of Gauge Theory,}
  JHEP {\bf 1003} (2010) 056
  [arXiv:0812.1840 [hep-th]].

\bibitem{Lada} T.~Lada,

{\it $L_\infty$ algebra representations},
Applied Categorical Structures {\bf 12} (2004) 29-34.

\bibitem{Stasheff1} J. Stasheff, {\it unpublished}.

\bibitem{FLS} R.Fulp, T.Lada and J.Stasheff, {\it Sh-Lie algebras induced by
gauge transformations}.


 
\bibitem{Sakharov}
  A.~D.~Sakharov,
 {\it Vacuum quantum fluctuations in curved space and the theory of
gravitation,}
  Sov.\ Phys.\ Dokl.\  {\bf 12} (1968) 1040
   [Dokl.\ Akad.\ Nauk Ser.\ Fiz.\  {\bf 177} (1967) 70]
   [Sov.\ Phys.\ Usp.\  {\bf 34} (1991) 394]
   [Gen.\ Rel.\ Grav.\  {\bf 32} (2000) 365]. 

\bibitem{BCLPS} L.~Bonora, M.~Cvitan, P.~Dominis Prester, B.~Lima de Souza and
I.~Smoli\'{c},
  {\it Massive fermion model in 3d and higher spin currents,}
  JHEP {\bf 1605} (2016) 072
    [arXiv:1602.07178 [hep-th]].

 \bibitem{BCDGLS} L.~Bonora, M.~Cvitan, P.~Dominis Prester, S.~Giaccari, B.~Lima
de Souza and T.~\v{S}temberga,
{\it One-loop effective actions and higher spins,}
  JHEP {\bf 1612} (2016) 084
  [arXiv:1609.02088 [hep-th]].

 \bibitem{BCDGS} L.~Bonora, M.~Cvitan, P.~Dominis Prester, S.~Giaccari,  and
T.~\v{S}temberga,
{\it One-loop effective actions and higher spins. II},
  JHEP {\bf 1801} (2018) 080 
  [arXiv:1709.01738 [hep-th]].

 
\bibitem{Strassler}
  M.~J.~Strassler,
 {\it Field theory without Feynman diagrams: One loop effective actions,}
  Nucl.\ Phys.\ B {\bf 385} (1992) 145
  [hep-ph/9205205].

\bibitem{Segal}
  A.~Y.~Segal,
{\it Conformal higher spin theory,}
  Nucl.\ Phys.\ B {\bf 664} (2003) 59
  [hep-th/0207212].

A.~Y.~Segal, 
{\it Point particle in general background fields vsersus gauge theories of
traceless symmetric tensors,}
  Int.\ J.\ Mod.\ Phys.\ A {\bf 18} (2003) 4999
  [hep-th/0110056].


\bibitem{Schmidt}
  M.~G.~Schmidt and C.~Schubert,
 {\it The Worldline path integral approach to Feynman graphs,}
  [hep-ph/9412358].

 M.~G.~Schmidt and C.~Schubert,
{\it Worldline Green functions for multiloop diagrams,}
  Phys.\ Lett.\ B {\bf 331} (1994) 69
  [hep-th/9403158].

 \bibitem{Dai}
  P.~Dai and W.~Siegel,
 {\it Worldline Green Functions for Arbitrary Feynman Diagrams,}
  Nucl.\ Phys.\ B {\bf 770} (2007) 107
  [hep-th/0608062].

\bibitem{Bonezzi}
  R.~Bonezzi,
{\it Induced Action for Conformal Higher Spins from Worldline Path Integrals,}
  Universe {\bf 3} (2017) no.3,  64
  [arXiv:1709.00850 [hep-th]].

{
\bibitem{BCDGSII} L.~Bonora, M.~Cvitan, P.~Dominis Prester, S.~Giaccari,  and
T.~\v{S}temberga,
{\it HS in flat spacetime. YM-like models},    [arXiv:1812.05030 [hep-th]].
}

{
\bibitem{Lopatin} 
  V.~E.~Lopatin and M.~A.~Vasiliev,
 {\it Free Massless Bosonic Fields of Arbitrary Spin in $d$-dimensional De 
Sitter Space,}
  Mod.\ Phys.\ Lett.\ A {\bf 3}, 257 (1988).
  doi:10.1142/S0217732388000313}

\bibitem{Bekaert}
  X.~Bekaert, N.~Boulanger and P.~Sundell,
{\it How higher-spin gravity surpasses the spin two barrier: no-go theorems 
versus yes-go examples,}
  Rev.\ Mod.\ Phys.\  {\bf 84} (2012) 987
  [arXiv:1007.0435 [hep-th]].
 

\bibitem{BoosDavy} E. E. Boos and Andrei I. Davydychev, {\it A Method of
evaluating massive Feynman integrals}, Theor.\
Math.\ Phys. {\bf 89}  (1991) 1052 [Teor. Mat. Fiz.89,56(1991)].

Andrei I. Davydychev, {\it A Simple formula for reducing Feynman diagrams to
scalar integrals}, Phys.\ Lett.\ {\bf B263} (1991) 107.

Andrei I. Davydychev, {\it Recursive algorithm of evaluating vertex type Feynman
integrals}, J.\ Phys.\ A25 (1992) 5587.
 


\end{thebibliography}
\end{document}